\documentclass[12pt]{iopart}
\usepackage{graphicx}
\usepackage{hyperref}
\usepackage{multirow}

\begin{document}

\title{Fundamental studies of superconductors using scanning magnetic imaging}

\author{J.R. Kirtley}

\address{Center for Probing the Nanoscale, Stanford University, Stanford, California, USA}

\begin{abstract}
In this review I discuss the application of scanning magnetic imaging to fundamental studies of superconductors,  concentrating on three scanning magnetic microscopies - scanning SQUID microscopy (SSM), scanning Hall bar microscopy (SHM), and magnetic force microscopy (MFM).  I briefly discuss the history, sensitivity, spatial resolution, invasiveness, and potential future developments of each technique. I then discuss a selection of applications of these microscopies. I start with static imaging of magnetic flux:
An SSM study provides deeper understanding of vortex trapping in narrow strips, which are used to reduce noise in superconducting circuitry. 
Studies of vortex trapping in wire lattices, clusters, and arrays of rings and nanoholes show fascinating ordering effects. 
The cuprate high-T$_c$ superconductors are shown to have predominantly $d$-wave pairing symmetry by magnetic imaging of the half-integer flux quantum effect. 
Arrays of superconducting rings act as a physical analog for the Ising spin model, with the half-integer flux quantum effect helping to eliminate one source of disorder in antiferromagnetic arrangements of the ring moments.
Tests of the interlayer tunneling model  show that the condensation energy available from this mechanism can not account for the high critical temperatures observed in the cuprates. 
The strong divergence in the magnetic fields of Pearl vortices allows them to be imaged using SSM, even for penetration depths of a millimeter.  
Unusual vortex arrangements occur in samples comparable in size to the coherence length.
Spontaneous magnetization is not observed in Sr$_2$RuO$_4$, which is believed to have $p_x\pm ip_y$ pairing symmetry, although effects hundreds of times bigger than the sensitivity limits had been predicted. However, unusual flux trapping is observed in this superconductor.
Finally, unusual flux arrangements are also observed in magnetic superconductors. 
I then turn to vortex dynamics:
Imaging of vortices in rings of highly underdoped cuprates places limits on spin-charge separation in these materials.
Studies of spontaneous generation of fluxoids upon cooling rings through the superconducting transition provide clues to dynamical processes relevant to the early development of the universe, while
studies of vortex motion in cuprate grain boundaries allow the measurement of current-voltage characteristics at the femtovolt scale for these technologically important defects. 
Scanning SQUID susceptometry allows the measurement of superconducting fluctuations on samples comparable in size to the coherence length, reveal 
stripes in susceptibility believed to be associated with enhanced superfluid density on the twin boundaries in the pnictide superconductor Co-doped Ba-122, and  indicate the presence of spin-like excitations, which may be a source of noise in superconducting devices, in a wide variety of materials.
Scanning magnetic microscopies allow
the absolute value of penetration depths to be measured locally over a wide temperature range, providing clues to the symmetry of the order parameter in unconventional superconductors. 
Finally, MFM tips can be used to manipulate vortices, providing information on flux trapping in superconductors. 

\end{abstract}

\tableofcontents
\newpage
\pagestyle{plain}

\section{Introduction}

Superconductivity, the complete absence of electrical resistance in some metals below a  critical temperature T$_c$, is one of the best understood phenomena in solid-state physics. As described in the Bardeen-Cooper-Schrieffer (BCS) theory \cite{bardeen1957mts}, superconductivity occurs in conventional superconductors through phonon-mediated pair-wise interaction of charge carriers to form Cooper pairs. The characteristic energy needed to form a single particle excitation from the superconducting condensate is the energy gap $\Delta$.  The Cooper pairs form a macroscopic quantum state characterized by a wave function $\psi(\vec{r})= | \psi(\vec{r}) | e^{i \phi(\vec{r})}$, where $\phi$ is the Cooper pair phase, with the density of Cooper pairs $n_s(\vec{r}) = | \psi(\vec{r})|^2$. The macrosopic quantum nature of the pairing state leads to a wide variety of behaviors: persistent currents, Meissner screening, flux quantization, the Josephson effects, and superconducting quantum interference. Superconductivity has become a multi-billion dollar industry, with applications in medical imaging, cell phone filters, electrical motors, power transmission, transportation, magnetic manipulation, magnetic sensing, etc.

However, there is still much to learn about superconductivity: Although superconductivity in the copper-oxide based perovskites \cite{bednorz1986pht,wu1987san} was discovered in 1986, despite almost 25 years of intense effort a consensus on the pairing mechanism in the cuprates has yet to emerge. The heavy Fermion superconductors \cite{steglich1979prl,stewart1984rmp} have large carrier masses, strong interaction between spin and charge degrees of freedom, and potentially a wide variety of Cooper pairing symmetries. The non-cuprate perovskite superconductor Sr$_2$RuO$_4$ \cite{maeno1994slp} is believed to have a $p$-wave pairing state that breaks time reversal symmetry \cite{mackenzie2003ssp}. Interest in unconventional superconductors has been reawakened with the discovery of iron-based compounds with high critical temperatures \cite{takahashi2008saf}. 

Scanning magnetic microscopies have played, and will continue to play, an important part in our efforts to understand superconductivity. There are now a large number of scanning magnetic microscopies. Although it is beyond the scope of this review to cover all of these techniques in detail, it is appropriate to briefly describe some,  and provide a few citations as an introduction to the literature: 

In Lorentz microscopy \cite{harada1992rto,harada1993vcd,matsuda1996odi,harada1996dov}, a coherent beam of high energy (300keV) electrons are passed through a thin sample and then imaged slightly off the focal plane. Magnetic fields introduce small phase shifts which cause interference in the defocussed image. Lorentz microscopy can image vortices with deep sub-micron spatial resolution at video rates and is sensitive not only to surface fields but also to fields within the sample. However, the sample must be thinned.  

In magneto-optic microscopy \cite{koblischka1995moi,larbalestier2001slc,golubchik2009cbh}, the sample is illuminated with linearly polarized light. It passes through a magneto-optically active layer, reflects off the sample surface, passes again through the magneto-optically active layer, and then through a crossed polarizer. Regions of the sample with magnetic fields produce rotations of the polarization axis of the light, which is detected as a bright spot. Magneto-optic microscopy has the advantage of relative simplicity, but has limited spatial resolution and sensitivity. Nevertheless, observation of individual superconducting vortices with 0.8$\mu m$ spatial resolution has recently been reported \cite{golubchik2009cbh}.

 In Sagnac interferometry \cite{kapitulnik1994hrm,petersen1998kri,xia2008pke}, a single beam of light is split into two components which travel along identical paths in opposite directions around a loop. Any effects which break the time reversal symmetry, including the magneto-optical effect, will cause interference, which can be sensitively detected. High spatial resolution can be achieved using scanning near-field optical microscopy \cite{petersen1998kri}.  

In spin polarized scanning tunneling microscopy \cite{wei1999tss,wachowiak2002doi,wiebe2004300} a scanning tunneling microscope has a tunneling tip with a ferromagnetic or antiferromagnetic atom from which the tunneling electrons emerge. The tunneling probability depends on the spin polarization of the sample, and therefore contrast is achieved if there are spatial variations in the atomic spin polarizations. Spin polarized tunneling is useful for studying superconductivity in, for example, the high-temperature cuprate superconductors, since it is believed that magnetism and superconductivity are intimately related in these materials. However, the spin polarized tunneling current is not directly related to the local magnetic fields. 

In diamond nitrogen-vacancy microscopy \cite{lifshitz1998odm,chernobrod2005smb,degen2008smf,balasubramanian2008nim,maze2008nms}, a magnetic field dependent shift in the energy levels of a single nitrogen-vacancy center in diamond is detected optically. There are two basic schemes for detecting this energy shift. In the ``dc" technique, microwaves at a fixed frequency as well as light are incident on the center. The intensity of the resultant fluorescence depends on the local magnetic field. This method can detect milliTesla fields with, in principle, nanometer scale spatial resolution. In the ``ac" technique, spin-echo microwave techniques are used to detect the modulation of the center's energy levels by the local magnetic field. The ``ac" technique has a sensitivity of 60 nT, with in principle a spatial resolution of a few nm, but has reduced sensitivity at zero frequency. These techniques have the advantages of good sensitivity, small invasiveness, room temperature operation, inherently high speed, and potentially high spatial resolution. It remains to be seen whether single nitrogen-vacancies in nm sized diamond crystallites can be placed at the end of scanning tips while still retaining long lifetimes. This is necessary to achieve high spatial resolution and sensitivity simultaneously.

A magnetic tunnel junction \cite{gallagher1997mmt,schrag2004cdm,schrag2006mci} is a planar structure with two ferromagnetic electrodes. The tunneling current through the junction depends on the relative alignment of the magnetic moments of the electrodes, and therefore depends on the magnetic field environment. Magnetic tunnel junctions suitable for scanning microscopy are commercially available. They currently have lateral dimensions (and therefore spatial resolutions) of about 4 microns, and field noises at 100 Hz of about 5$\times 10^{-8} T/Hz^{1/2}$ \cite{schrag2006mci}. They have the advantage of room temperature operation, but are not currently as sensitive as Hall bars or SQUIDs. 

%For example, I will review how scanning SQUID microscopy played a central role in the demonstration that the cuprate high temperature superconductors have predominantly $d$-wave pairing symmetry, and the role it is currently playing in attempting to understand the pnictide superconductors. In addition to shedding light on the fundamental mechanisms of superconductivity, scanning magnetic microscopies have also used superconducting systems as physical analogues for such diverse topics as the Ising spin system and the early evolution of the universe.

There are already excellent reviews covering the topic of magnetic imaging of superconductors \cite{hartmann1994spm,delozanne1999spm,bending1999lmp}, and there has been an enormous amount of work in this area. I  will therefore not attempt to survey the field exhaustively, but will only cover recent work in three scanning magnetic microscopy techniques: scanning SQUID, scanning Hall bar, and magnetic force microscopies. I will present a short introduction to the basics, advantages and shortcomings of each technique, and then present a selection of fundamental applications of these magnetic microscopies in the area of superconductivity. 
%This selection is heavily biased towards applications I have been involved in, and therefore am most familiar with. I therefore neglect much important work in the area. I apologize in advance for that.

\section{Fundamentals}

\subsection{Scanning SQUID microscopy}
Josephson \cite{josephson1962pne} predicted in 1962 the possibility of coherent tunneling of Cooper pairs between two weakly coupled superconductors, and wrote down the fundamental relations:
\begin{eqnarray}
I_s = I_0 \sin \varphi  \nonumber \\
V = \frac{\hbar}{2e} \frac{d\varphi}{dt},
\label{eq:josephson}
\end{eqnarray}
where $I_s$ is the supercurrent through the weak link,  $\varphi$ is the difference between the Cooper pair phases on the two sides of the weak link,  $V$ is the voltage across the weak link, and $I_0$ is the critical current, the maximum current that can pass through the weak link before a voltage develops. A physical Josephson junction can be modeled as an ideal Josephson element in parallel with a capacitor and a resistor in the resistively shunted junction (RSJ) model \cite{mccumber1968twl,stewart1968cvc}. This model obeys the  equation:
\begin{equation}
I_B = \frac{\hbar}{2eR} \frac{d \varphi}{dt} + I_0 \sin \varphi + \frac{\hbar C}{2e} \frac{d^2  \varphi}{dt^2}+ I_n \, ,
\label{eq:stewart_mccumber}
\end{equation}
where $I_B$ is the bias current through the weak link, $R$ is the resistance, $C$ is the capacitance, and $I_n$ is a noise current, given for the case of an ideal Johnson noise source resistor by 
\begin{equation}
<I_n>^2 = 4 k_B T \Delta f/R \, ,
\label{eq:johnson}
\end{equation} 
where $<I_n>^2$ is the time-averaged current noise squared, and $\Delta f$ is the frequency band width over which the current noise is measured. This equation has the form of the equation of motion of a driven, damped harmonic oscillator. The dynamics of a Josephson weak-link can  be visualized as that of a particle moving in a sinusoidal ``washboard" potential with oscillation amplitude $I_0 \Phi_0/2\pi$, where $\Phi_0=h/2e$ is the superconducting flux quantum, and the average slope of the potential is given by $<dE/d \varphi>=\Phi_0(I_B-I_n)/2\pi$. In the absence of a noise current,  the particle will escape from a local potential minimum and run down the washboard potential if the bias current $I_B$ is greater than $I_0$, in the process developing a voltage given by the second of Eq.s \ref{eq:josephson}. The solutions to Eq. \ref{eq:stewart_mccumber} can be divided into two classes, according to whether the Stewart-McCumber parameter $\beta_c = 2\pi I_0 R^2 C/\Phi_0$ is greater  or less than 1. If $\beta_c < 1$,  a running particle retraps into a local potential minimum ($V=0$), as soon as $I_B = I_0$, and the current-voltage characteristic is non-hysteretic. On the other hand, if $\beta_c > 1$, a running particle is not retrapped until $I_B < I_0$, and the current-voltage characteristic is hysteretic. In the presence of noise, the SQUID can be thermally excited out of the zero-voltage state at bias current $I_B < I_0$ \cite{fulton1974lzv}. In addition, the weak-link can macroscopically quantum tunnel through the barrier from the zero-voltage to the voltage state \cite{voss1981mqc}.

\begin{figure}[tb]
        \begin{center}
                \includegraphics[width=10cm]{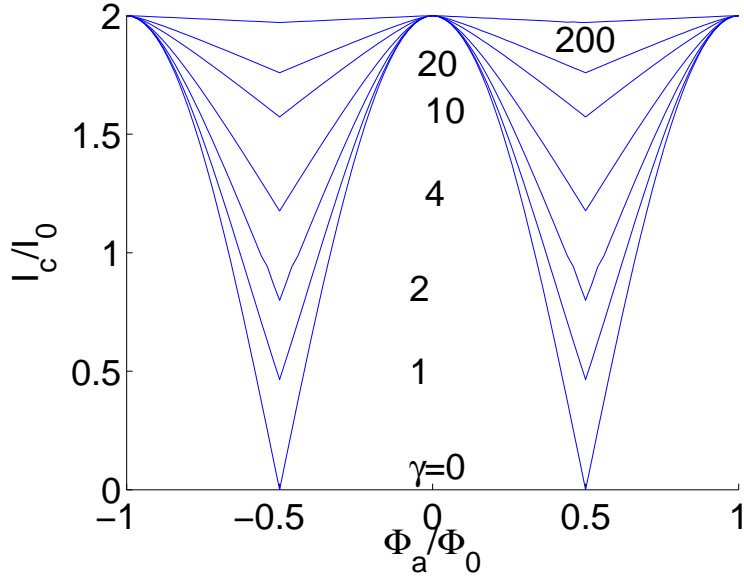}
        \end{center}
        \caption{Plot of the critical current $I_c$ of a symmetric (L$_1$ = L$_2$ = L/2, I $_1$=I$_2$=I$_0$) 2-weak link SQUID as a function of applied flux, for different values of the parameter $\gamma=2\pi L I_0/\Phi_0$. I$_0$ is the critical current of one of the weak-links, and $L$ is the total inductance of  the SQUID.}
        \label{fig:sqicphi}
\end{figure}

A Superconducting Quantum Interference Device (SQUID) is a superconducting ring interrupted by at least one Josephson weak link. For SQUID microscope applications a  SQUID with two weak-links is most often used. Such a SQUID is labelled a dc-SQUID, since the modulation of the SQUID critical current with applied magnetic field can be observed using time-independent bias currents. Consider a SQUID with two weak-links $1,2$ with critical currents $I_{c1}, I_{c2}$, phase drops across the weak-links $\varphi_1, \varphi_2$, and inductances in the two arms of the SQUID $L_1, L_2$. Neglecting for the moment the current through the resistive and capacitive elements of the junctions, the bias current across the SQUID $I_B$ and the magnetic flux through the SQUID $\Phi$ are given by
\begin{eqnarray}
I_B=I_1 \sin \varphi_1 + I_2 \sin \varphi_2 \nonumber \\ 
\Phi = \Phi_a - L_1 I_1 + L_2 I_2, 
\label{eq:squid_sum_conditions}
\end{eqnarray}
where $\Phi_a$ is the externally applied flux. Integrating all the changes in phase around the SQUID loop along a path sufficiently deep inside the superconductor that there are no supercurrents along the integration path (i.e. they are shielded out by the bulk of the superconductor), and using the fact that the canonical momentum of the Cooper pairs is given by ${\vec p}_{\rm canonical} = \vec{p}+q \vec{A}$, where $q=2e$ is the charge on the Cooper pairs and $\vec{A}$ is the vector potential, and finally by insisting that the Cooper pair wavefunction be single valued, leads to 
\begin{equation} 
2 \pi n = \varphi_2 - \varphi_1+ \frac{2 \pi}{\Phi_0} (\Phi_a + L_2 I_2 - L_1 I_1) \,
\label{eq:flux_quantization_condition}
\end{equation}
$n$ an integer. Solutions to Equations \ref{eq:squid_sum_conditions} and \ref{eq:flux_quantization_condition} are shown in Figure \ref{fig:sqicphi} for the case of symmetric weak-link critical currents ($I_{c1} = I_{c2}$) and arm inductances ($L_1 = L_2$). The parameter $\gamma=2\pi L I_0/\Phi_0$ (L=L$_1$+L$_2$) determines the depth of the modulation of the SQUID critical current with flux: larger SQUID inductances lead to smaller modulations. The solutions for non-symmetric SQUIDs are more complicated, but qualitatively similar \cite{smilde2004psb} to those shown here.

There are 4 relevant sources of noise in dc SQUIDs: Johnson noise arises from thermal fluctuations, assumed to be dominated by the shunt resistor in the RSJ model \cite{tesche1977dcs}; shot noise arises from the discrete charge of the quasiparticles \cite{dahm1969lre} and Cooper pairs \cite{stephen1968tjo} traversing the weak-links; quantum noise arises from zero point motion in the shunt resistors \cite{koch1980qnt}; and $1/f$ noise can arise from a number of sources. Tesche and Clarke calculate \cite{tesche1977dcs} that the signal to noise ratio in dc SQUIDs is optimized when $\beta_L=2LI_0/\Phi_0 \approx 1$. In addition, the Stewart-McCumber parameter $\beta_c$ should be about 1, so that the SQUID is just non-hysteretic. If these two conditions are met, the optimal flux noise power $S_{\Phi}$ from the first three sources is given by:
%\begin{eqnarray}
%                              & = 4 k_B T L (\pi L C)^{1/2} &\,\,{\rm Johnson \, noise} \nonumber \\
%S_{\Phi}                 & = hL                                   &\,\,{\rm Shot \, noise} \nonumber \\                 
%                             & =\hbar L                                 &\,\,{\rm Quantum \, noise} 
%\label{eq:squid_noise}
%\end{eqnarray}
\begin{equation}
S_{\Phi} = \left\{
\begin{array}{ll}
 4 k_B T L (\pi L C)^{1/2} &{\rm Johnson \, noise} \nonumber \\
    hL                                   &{\rm Shot \, noise} \nonumber \\ 
 \hbar L                                 &{\rm Quantum \, noise}.
\end{array} \right.
\end{equation}
Typical values for scanning SQUID sensors using a Nb-Al$_2$O$_3$-Nb trilayer process are $L$ = 100 pH, I$_0$ = 6 $\mu$A, C = 8.36 $\times$ 10$^{-13}$ F, R = 5 $\Omega$ and T = 4.2 K. These parameters lead to $\Phi_{\rm Johnson} = \sqrt{S_{\Phi}}$ = 3 $\times 10^{-7} \Phi_0/\sqrt{\rm Hz}$ for the flux noise due to Johnson noise; $\Phi_{\rm shot} = 1.2 \times 10^{-7} \Phi_0 / \sqrt{\rm Hz}$ for shot noise; and $\Phi_{\rm quantum} = 5 \times 10^{-8} \Phi_0/ \sqrt{Hz}$ for noise due to zero point motion. 

In the frequency range of interest Johnson noise, shot noise, and quantum noise are ``white": the frequency distribution of the noise is independent of frequency $f$. $1/f$ noise, as its name implies, has a noise amplitude that follows a $1/f$ dependence. It almost always dominates the noise in SQUIDs at sufficiently low frequencies. In the Dutta, Dimon, and Horn model \cite{dutta1979esn} $1/f$ noise results from the superposition of ``shot-like" contributions from two-level traps with a large distribution of trap energies. These traps could be, for example, defects in the oxide in a tunnel junction, or trap sites for superconducting vortices in the superconducting bulk making up the SQUID. Koch, Divincenzo, and Clarke \cite{koch2007mff} have suggested that $1/f$ noise in SQUIDs could result from coupling of flux into the SQUID from electronic spins with a wide distribution of characteristic times. 

As can be seen in Figure \ref{fig:sqicphi}, the critical current of the SQUID is periodic in flux, with a period given by $\Phi_0$. The most commonly used scheme for determining this magnetic flux uses $ac$ flux modulation in a flux locked loop \cite{clarke1989pas,kirtley1999ssm}. If the SQUID is current biased just above the critical current the voltage varies sinusoidally with flux. If the $dc$ component of the total (external+feedback) flux  is such that the SQUID voltage is at an extremum, the first harmonic voltage response will be zero and any flux change will give a linear response. Negative feedback on the $dc$ component of the flux is then used to keep the response zeroed: this feedback signal is linearly proportional to the flux threading the SQUID, with a constant of proportionality that can  easily be determined with great precision. Traditionally the voltage signal from the SQUID has been amplified at low temperatures using a tuned $LC$ circuit or a transformer, and at room temperature using phase sensitive lock-in detection. However, SQUID array amplifiers can also be used \cite{welty1991sad,huber2001dcs}. These have the advantage of bandwidths of  over 100 MHz, and no requirement for $ac$ flux modulation.

%\subsection{Scanning with SQUIDs}

For this review I will concentrate on high-spatial resolution magnetic microscopies. High spatial resolution is achieved by placing a small sensor close to the sample. In the case of SQUID microscopy this means making the SQUID small, or making a pickup loop integrated into the SQUID small. The scaling of the flux signal and spatial resolution with sensor size depends on the type of field source. For this review I will concentrate on three basic field sources: A point dipole $\vec{m}$ generates the magnetic induction
\begin{equation}
\vec{B}=\frac{\mu_0}{4\pi}\frac{3 \hat{r}(\hat{r}\cdot \vec{m})-\vec{m}}{r^3}, \hspace{1in} \rm{Dipole \, field}
\label{eq:dipole_field}
\end{equation}
where $\hat{r}$ is the direction and $r$ is the magnitude of the vector between the dipole source and the point of interest. The magnetic flux $\Phi_s$ through a square area of side $d$ oriented parallel to the $xy$ plane and centered at $\vec{r}=z\hat{z}$ above a dipole at the origin with moment $m$ oriented parallel to the $z$ axis is given by
%\begin{equation}
%\Phi_s = \frac{\sqrt{2}\mu_0md}{\pi}\left \{  \sqrt{ d^2-2dx_0+2(xo^2+z0^2)}\left[-4d^2x_0(x_0^2-2z^2)+16x_0(x_0^2+z^2)(x_0^2+2z^2)+d^5-2d^3(3x_0^2-4z^2)+8d
%(x_0^2+x_0^2z^2+2z^4)\right] \right . \\  \left .\sqrt{d^2+2dx_0+2(x_0^2+z^2)}\left[+4d^2x_0(x_0^2-2z^2)-16x_0(x_0^2+z^2)(x_0^2+2z^2)-2w^3(3x_0^2-4z^2)+8d
%(x_0^4+x_0^2z^2+2z^4)\right] \right \} /\\ (d^2+4z^2)((d-2x_0)^2+4z^2)((d+2x_0)^2+4z^2)\sqrt{d^2+4d^2z^2+4(x_0^2+z^2)^2}
%\label{eq:dipole_field_full}
%\end{equation}
\begin{equation}
\Phi_s=\frac{2\mu_0md^2}{\pi \sqrt{d^2/2+z^2}(d^2+4 z^2)}. \hspace{0.58in} {\rm Dipole \, flux}
\label{eq:dipole_flux}
\end{equation}
In the limit $z\rightarrow 0$ $\Phi_s \rightarrow 2\sqrt{2}\mu_0m/\pi d$.

A superconducting vortex generates the induction
\begin{equation}
\vec{B}=\frac{\Phi_0}{2\pi r^2} \hat{r}, \hspace{1.8in} \rm{Monopole \, field}
\label{eq:monopole_field}
\end{equation}
where $\Phi_0=h/2e$ is the total magnetic flux generated by the vortex. The magnetic flux $\Phi_s$ through a square area of side $d$ oriented parallel to the $xy$ plane and centered at $\vec{r}=z\hat{z}$ above a vortex at the origin is given by 
\begin{equation}
\Phi_s=\frac{2\Phi_0}{\pi } {\rm tan^{-1}} \left ( \frac{d^2}{2z\sqrt{2d^2+4z^2}} \right ). \hspace{0.3in} {\rm Monopole \, flux}
\label{eq:monopole_flux}
\end{equation}
In the limit $z \rightarrow 0$ $\Phi_s \rightarrow \Phi_0$. Finally, an infinitely long and narrow line of current $I$ in the $\hat{y}$ direction at $x=0, z=0$ generates the field
\begin{equation}
\vec{B}=\frac{\mu_0 I}{2\pi r} \hat{\theta}, \hspace{2.1in} \rm{Current \, field}
\label{eq:current_field}
\end{equation}
where $\hat{\theta} = (x\hat{z}-z\hat{x})/r$. The magnetic flux $\Phi_s$ through a square area of side $d$ oriented parallel to the $xy$ plane at a height $z$ above this line has peaks at $x=\pm \sqrt{d^2+z^2}/2$ with magnitude 
\begin{equation}
\Phi_s=\frac{\mu_0Id}{4\pi} {\rm ln} \left [ \frac{(d-\sqrt{d^2+z^2})^2+4z^2}{(d+\sqrt{d^2+z^2})^2+4z^2} \right ] \hspace{0.3in} {\rm Current \, flux}
\label{eq:current_flux}
\end{equation}
In the limit $z \rightarrow 0$ the magnetic fluxes at the peaks diverge logarithmically. 

\begin{figure}[tb]
        \begin{center}
                \includegraphics[width=14cm]{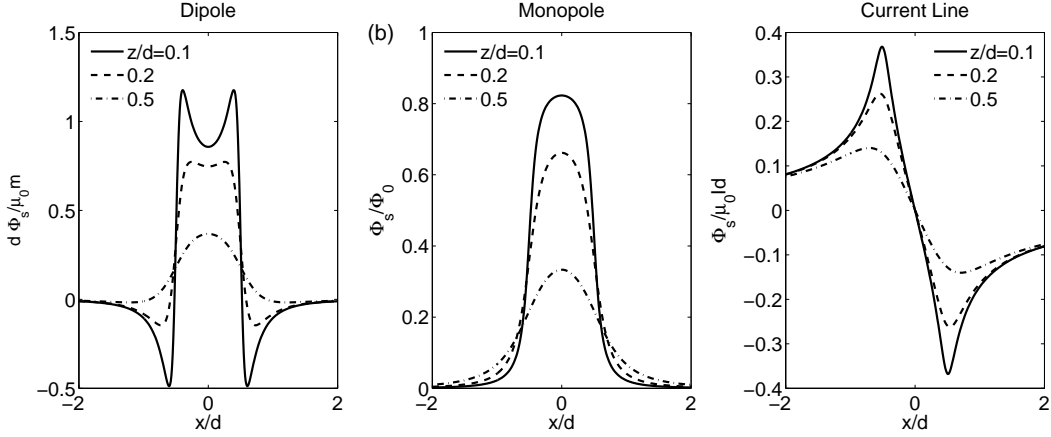}
        \end{center}\caption{Plot of the flux through a square area of side $d$ a height $z$ above a (a) dipole source with moment $m$ oriented normal to the scan plane (b) a monopole source with flux $\Phi_0$, and (c) a line source with current $I$. For a given value of $z/d$ the flux signal scales like $d^{-1}$ for a dipole source, is independent of $d$ for a monopole source, and scales like $d$ for a current line source.}
                \label{fig:threesrc}
\end{figure}

Figure \ref{fig:threesrc} plots the calculated flux through a square area of side $d$, as a function of $x$, the lateral position of the SQUID relative to the field source, for various spacings $z$ between pickup loop and sample, for these three different sources of field. For a given value of $z/d$ for a dipole source the flux signal through the SQUID gets larger as $d$ gets smaller, scaling like $1/d$. For a monopole source, such as a superconducting vortex, the peak flux is independent of $d$ at constant $z/d$. For a line of current, the peak flux signal increases like $d$ as the pickup loop diameter increases. In all cases the spatial resolution is about $d$ for $z<<d$.

\begin{figure}[tb]
        \begin{center}
                \includegraphics[width=10cm]{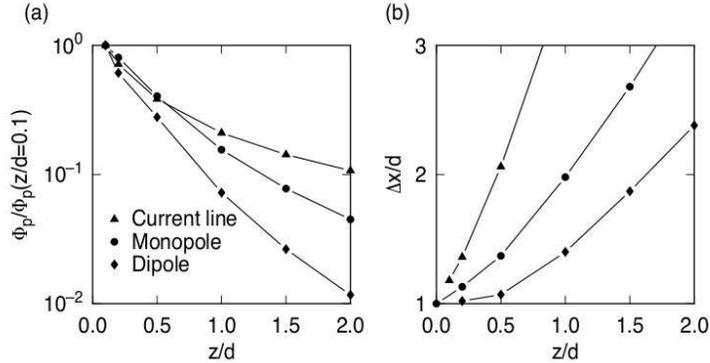}
        \end{center}
        \caption{(a) Plot of the ratio of the flux through a square area of side $d$ at a height $z$ to that at a height $z/d=0.1$ for various sources of field. (b) Full width at half-maximum for the dipole and monopole field sources, and the peak-to-peak distance for the current line source, of the flux through a square area of side $d$ as a function of height $z$.}
        
        \label{fig:squid_flux_and_width_vs_z}
\end{figure}

Figure \ref{fig:squid_flux_and_width_vs_z}a plots the ratio of the peak flux at a sensor height $z$ to the peak flux at $z/d=0.1$ for the three different field sources, while Figure \ref{fig:squid_flux_and_width_vs_z}b plots the full width at half-maximum $\Delta x$ for the dipole and monopole field sources, and the peak-to-peak distance for the current line source, as a function of sensor height $z$. The peak amplitude falls off nearly exponentially with sensor height for a dipole source, but falls off less quickly for a monopole or current line source. Similarly, the spatial resolution falls off quickly with sensor height for the dipole source, but less quickly for the monopole and current line sources. 

The same qualitative conclusions can be made for more complex sources of field {\cite{roth1989umi}. Consider a geometry in which the sample takes up a half-space $z<0$. For the static case the magnetic fields in the half space $z>0$ can be written as the gradient of a scalar potential $\varphi_m$: $\vec{B}=\vec{\bigtriangledown} \varphi_m$, where $\bigtriangledown^2 \varphi_m=0$. If we resolve the fields at the surface of the sample into Fourier components
\begin{equation}
\vec{b}_k(z=0) = \frac{1}{(2\pi)^2} \int_{-\infty}^{\infty} \int_{-\infty}^{\infty} \vec{B}(x,y,z=0) e^{i(k_x x + k_y y)} dx\, dy, 
\end{equation}
each of the components of the field for $z>0$ will be of the form $\vec{b}_k(z) = \vec{b}_k(z=0)e^{-k z}$, where $k=\sqrt{k_x^2+k_y^2}$: The fields will decay exponentially as $z$ gets larger, with the higher Fourier components decaying more rapidly. Therefore to get optimal sensitivity and spatial resolution, the sensor must be placed close to the sample.

It was recognized almost immediately after the demonstration of the Josephson effect \cite{anderson1963poj} and superconducting quantum interference effects  \cite{jaklevic1964qie} that local magnetic fields could be imaged by scanning samples and SQUIDs relative to each other \cite{zimmerman1964qfp}. However, it was not until 1983 that the first two-dimensional scanning SQUID microscope was built by Rogers and Bermon at IBM Research \cite{rogers1983deo}. This instrument was used to image superconducting vortices in association with the IBM Josephson computer program. Other notable early efforts in SQUID microscopy were by the Wellstood group at the University of Maryland \cite{mathai1993odm,black1993mmu,black1995irf}, the van Harlingen group at the University of Illinois \cite{vu1993dis,vu1993imv},  the Wikswo group at Vanderbilt University \cite{ma1993hri}, the Clarke group at U.C. Berkeley \cite{chemla2000umb} and the Kirtley group at IBM Research \cite{tsuei1994psf,kirtley1995hrs}. Kirtley and Wikswo \cite{kirtley1999ssm} reviewed some of fundamentals and early work concerning scanning SQUID microscopy.

There are several competing strategies for achieving good spatial resolution in a SQUID microscope sensor. The first strategy is to make the SQUID very small ($\mu$-SQUID), with narrow and thin constrictions for the Josephson weak links \cite{wernsdorfer1997een,veauvy2002smus}.  This strategy has the advantage of simplicity, since only one level of lithography is required. The original $\mu$-SQUIDs had hysteretic current-voltage characteristics. This meant that flux-locked-loop feedback schemes could not be used, and the flux sensitivity of these SQUIDs was relatively poor.  However, recently non-hysteretic, sub-micron sized SQUIDs have been made \cite{troeman2007nbn,granata2008ism,hao2008man}. Hao {\it et. al} \cite{hao2008man} made their SQUIDs non-hysteretic by using a second layer of tungsten to shunt them, and reported white noise floor levels of 0.2$\mu \Phi_0/{\rm Hz}^{1/2}$ in a SQUID with a diameter of about 370 nm. 

\begin{figure}[tb]
        \begin{center}
                \includegraphics[width=10cm]{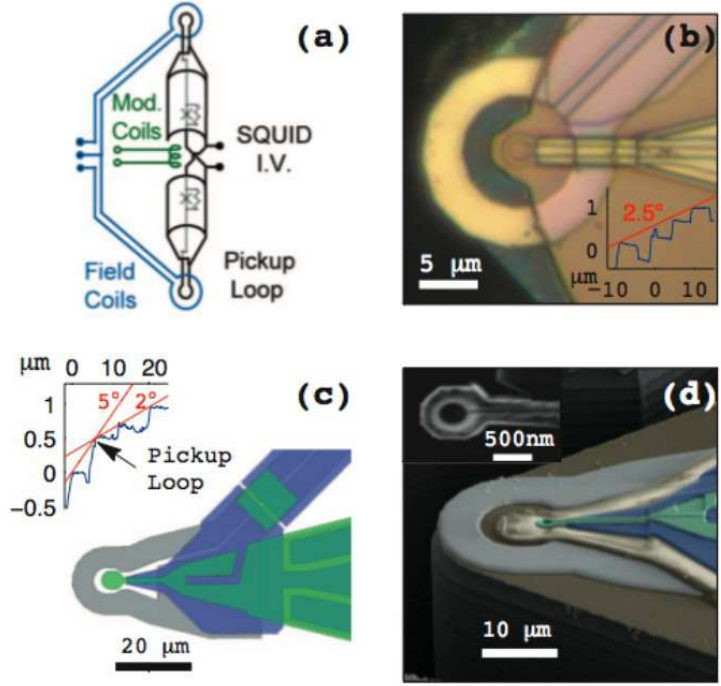}
        \end{center}
        \caption{Advanced integrated scanning SQUID susceptometer design. (a) Diagram of a counterwound SQUID susceptometer. (b) Optical micrograph of pickup loop region of an optically defined SQUID susceptometer. The outer ring is the field coil and the inner ring is the flux pickup loop of the SQUID. The edge of the silicon substrate is just visible on the left side. The inset shows an atomic force microscopy cross-section of the structure along a horizontal line through the center of the pickup loop. The pickup loop is closest to the sample when the tip is aligned at a 2.5$^o$ angle.  (c) Design for a SQUID sensor in which the pickup loop is defined using focussed ion beam (FIB) etching. In this case the pickup loop can touch down first if the alignment angle is between 2$^o$ and 5$^o$. (d) Scanning electron microscopy images of the sensor after FIB definition of the pickup loop. Reprinted figure with permission from N.C. Koshnick, M.E. Huber, J.A. Bert, C.W. Hicks, J. Large, H. Edwards and K.A. Moler, \href{http://apl.aip.org/applab/v93/i24/p243101_s1}{Appl. Phys. Lett. {\bf 93}, 243101 (2008)}. Copyright 2008 by the American Institute of Physics.}

        \label{fig:koshnick_squid}
\end{figure}

A second strategy is a self-aligned SQUID recently reported by Finkler {\it et al.} \cite{finkler2010san}. In this work three aluminum evaporations are made onto a quartz tube that has been pulled into a sharp tip with apex diameter between 100 and 400 nm. The first two evaporations, performed at an angle of 100 degrees relative to the axis of the tube, form the leads. A third evaporation along the tip axis forms a ring with two weak links to the leads, forming a dc SQUID. Finkler {\it et al.} report non-hysteretic current-voltage characteristics and a flux sensitivity of 1.8$\times 10^{-6} \Phi_0$/Hz$^{1/2}$ for SQUIDs with an effective area of 0.34 $\mu$m$^2$, operating at fields up to 0.6 T. 

A third strategy is to make a more conventional SQUID, but to have well shielded superconducting leads to a small pickup loop integrated into it \cite{ketchen1989dfa,ketchen1995dap,kirtley1995hrs} (see Fig. \ref{fig:koshnick_squid}). This has the disadvantage of complexity, since multiple levels of metal are required to shield the pickup loop leads, but has the advantages of reduced interaction between the SQUID and the sample and ease of using flux feedback schemes. Figure \ref{fig:koshnick_squid} illustrates the properties of current advanced scanning SQUID sensor design using the integrated pickup loop strategy \cite{koshnick2008ats}. This SQUID sensor uses a Nb-AlO$_x$-Nb trilayer for the junctions,  two levels of Nb for wiring and shielding, and SiO$_2$ for insulation between the various levels, as well as local field coils integrated into the pickup loop region, to allow local magnetic susceptibility measurements \cite{ketchen1989dfa,gardner2001ssq}. High symmetry in the SQUID and field coil layout (see Fig. \ref{fig:koshnick_squid}a) and tapered terminations are desirable to discriminate against background fields and reduce unwanted electromagnetic resonances \cite{huber2008gms}. Fig. \ref{fig:koshnick_squid}b is an optical micrograph of a scanning SQUID susceptometer in which both the field coil and pickup loop were defined using optical lithography. Fig. \ref{fig:koshnick_squid}c shows the layout for a device in which the pickup loop area is left as a ``tab" in the optical lithography step, and then patterned into a loop using focussed ion beam lithography \cite{freitag2006eel}. The completed device is shown in a scanning electron micrograph in Fig. \ref{fig:koshnick_squid}d. This device uses a non-planarized process, meaning that each successive layer conforms to the topography of the previous one. This leads to constraints on the alignment angles that can be used to get the minimum spacing between pickup loop and sample, as illustrated in the insets in Fig. \ref{fig:koshnick_squid}b,c. These constraints can be reduced by using a planarized process \cite{ketchen1991smup}, in which the insulating layers are chemically-mechanically polished to planes before successive Nb wiring levels are added.

If we assume a SQUID or SQUID pickup loop diameter of 1$\mu$m, a SQUID-sample spacing of 0.1$\mu$m, and a SQUID noise level of 10$^{-6}\Phi_0$/Hz$^{1/2}$, and using the peak signals from Fig. \ref{fig:threesrc}, we find a minimum detectable dipole signal of 139 electron spins/Hz$^{1/2}$, a minimum flux signal of 1.25$\times 10^{-6}\Phi_0$/Hz$^{1/2}$, and a minimum detectable line of current signal of 1.65 nA/Hz$^{1/2}$.

\subsection{Hall bar microscopy}

\begin{figure}[tb]
        \begin{center}
                \includegraphics[width=10cm]{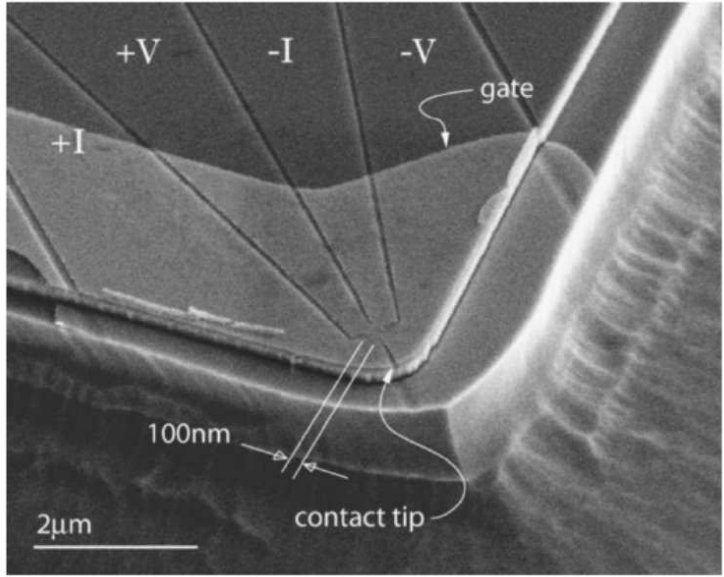}
        \end{center}
        \caption{Scanning electron microscope image of a 100 nm Hall probe. The four leads are separated by narrow etch lines. The gate shields the Hall cross from stray electrical charges and allows the modulation of the 2DEG beneath. The Hall probe will touch the sample surface at the contact tip. Reprinted figure with permission from C.W. Hicks, L. Luan, K.A. Moler, E. Zeldov, and H. Shtrikman,  \href{http://apl.aip.org/applab/v90/i13/p133512_s1}{Appl. Phys. Lett.  {\bf 90}, 133512 (2007)}. Copyright 2007 by the American Institute of Physics.}
        \label{fig:hicks_hallbar}
\end{figure}

A Hall bar develops a transverse voltage $V_{x}=-I_yB_z/n_{2d}e$ when a current $I_y$ passes through it in the presence of a magnetic induction $B_z$ perpendicular to the plane of the Hall bar \cite{hartmann1994spm,delozanne1999spm,bending1999lmp}, where $n_{2d}$ is the carrier density per unit area of the Hall bar. Therefore materials with small carrier densities such as semi-metals, semi-conductors, or two-dimensional electron gases at the interface between semi-conductors with different bandgaps  develop larger Hall voltages. Early scanning Hall bar systems used evaporated films of bismuth \cite{broom1962sis}, InSb \cite{weber1973mfd}, or GaAs \cite{tamegai1992doc}. More recently high sensitivity and spatial resolution have been achieved with GaAs/Al$_x$Ga$_{1-x}$As heterojunction structures \cite{chang1992shp,chang1992shp2,oral1996rts,oral1998dom,grigorenko2001shp,dinner2005csh}. In addition, 250 nm sized scanning probes using GaSb/InAs/GaSb \cite{grigorenko2001shp}, and micron-scale Si/SiGe and InGaAs/InP Hall crosses have been characterized at low temperatures \cite{vanVeen1999msh,cambel2000apr}. Hicks et al. \cite{hicks2007ncn} have fabricated Hall bars as small as 85 nm on a side (see Fig. \ref{fig:hicks_hallbar}) from GaAs/Al$_x$Ga$_{1-x}$As heterojunction electron gas material, and estimate a field noise of 500$\mu$T/Hz$^{1/2}$ and a spin sensitivity of 1.2$\times$10$^4\mu_B/Hz^{1/2}$ at 3 Hz and 9K for sensors 100 nm on a side. Sandhu {\it et al.} have fabricated 50 nm bismuth Hall bar sensors with a noise of 0.8G/$Hz^{1/2}$ \cite{sandhu2004fnh}.
At present Hall bars from GaAs/Al$_x$Ga$_{1-x}$As heterojunctions are less sensitive above about 100 K because of thermal excitations.  Scanning Hall bar sensors made of InAs have been used to image single magnetic biomolecular labels at room temperature  \cite{mihajlovic2007iqw}. 

Magnetic tunnel junctions and Hall bars in the ballistic regime \cite{peeters1998hmb,geim1997bhm,novoselov2003sph} produce a signal which is proportional to $<B_z>$, the magnetic field perpendicular to the sensor, averaged over the sensor. The situation is more complicated for a Hall bar in the diffusive regime \cite{bending1997heh}, but may be qualitatively similar. The average fields $<B_z>$ for the three sources of field can be inferred from Eq.s \ref{eq:dipole_flux}, \ref{eq:monopole_flux} and \ref{eq:current_flux}  (Fig. \ref{fig:threesrc}), by substituting $\Phi_s \rightarrow <B_z>d^2$ and in Fig. \ref{fig:squid_flux_and_width_vs_z} by substituting $\Phi_p \rightarrow <B_z>_p$, where $<B_z>_p$ is the peak value of the field averaged over the sensor area. For a given value of $z/d$, $<B_z>$ is proportional to $1/d^3$ for a dipole source, proportional to $1/d^2$ for a monopole source, and proportional to $1/d$ for a line of current source. Boero {\it et al.} \cite{boero2003mhd} estimate that the optimal Johnson noise limited minimum detectable magnetic field in a micro-Hall cross geometry is given by
\begin{equation}
B_{min} \approx \frac{\sqrt{4k_BTR_0}}{v_{sat}w}     \hspace{1in} {\rm Thermal}
\label{eq:boero}
\end{equation}
where $k_B$ is Boltzman's constant, $T$ is the operating temperature, $R_0$ is the output resistance at zero magnetic field, $v_{sat}$ is the saturation carrier drift velocity, and $w$ is the width of the cross. This results in $B_{min} \sim 2nT/\sqrt{Hz}$ for $w$=1$\mu$m at 300K for doped InSb. Thermal noise is expected to be independent of sensor size, and therefore the scaling for sensitivity is the same as for $<B_z>$. At the low temperatures used for imaging superconductors, transport in the Hall sensor is often quasi-ballistic and the signal-to-noise is better characterized by a mobility and an optimum probe current. The latter is typically defined as the maximum current before the onset of pronounced 1/f noise due to one of several possible sources, e.g., generation-recombination noise at deep traps or across heterointerfaces, heating or impact ionization. However, noise in micron scale Hall bars is often dominated by $1/f$ noise \cite{novoselov2003sph,hicks2007ncn}. The minimum detectable field can then be written as \cite{mihajlovic2007iqw}
\begin{equation}
B_{min} \approx \frac{1}{\mu d} \sqrt{\frac{\alpha_H G_N \Delta f}{n f}}, \hspace{0.7in} {\rm 1/f}
\label{eq:1/fnoise}
\end{equation}
where $\mu$ is the mobility, $d$ is the sensor size, $\alpha_H$ is Hooge's $1/f$ noise parameter, $G_N \sim 0.325$ is a constant, $n$ is the carrier density, $\Delta f$ is the measurement bandwidth, and $f$ is the frequency. Eq. \ref{eq:1/fnoise} combined with Eq.'s \ref{eq:dipole_flux}, \ref{eq:monopole_flux}, and \ref{eq:current_flux} imply that for a Hall bar dominated by $1/f$ noise, for a given $z/d$ the minimum detectable dipole moment scales like $d^2$, the minimum detectable flux scales like $d$, and the minimum detectable current is independent of $d$.  It has been reported that  low frequency noise in Hall bar devices can be significantly reduced by optimizing the voltage on a gate over the Hall cross \cite{li2004mns,hicks2007ncn} and that sufficiently small devices have noise which is composed of a single Lorentzian spectrum \cite{mueller2006don}. 
%Ultimately it should be possible to engineer scanning Hall bar sensors with spatial resolution below 100 nm and spin sensitivities below 1000 Bohr magnetons in a 1 Hz bandwidth.

\subsection{Magnetic force microscopy}

\label{sec:mfm_fundamentals}

\begin{figure}[tb]
        \begin{center}
                \includegraphics[width=16cm]{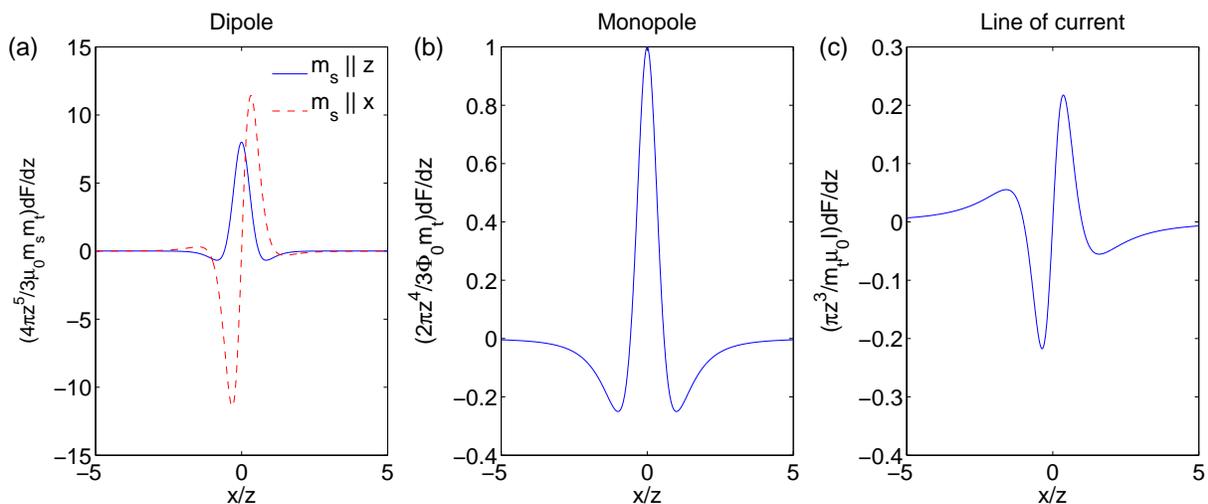}
        \end{center}
        \caption{MFM modeling using the point dipole tip approximation: Force derivative curves for (a) a dipole source located at the origin with magnetic moment $m_s$ oriented parallel to $z$ (solid line) or parallel to $x$ (dashed line), Eq. \ref{eq:dipole_field}, (b) a monopole source located at the origin with magnetic flux $\Phi_0$, Eq. \ref{eq:monopole_field}, and (c) a line of current $I$ along the $y$ axis, Eq. \ref{eq:current_field}. The tip is modeled as a point dipole with moment $m_{t}$ oriented parallel to the $z$-axis scanning a height $z$ above the $xy$ plane along the $x$ axis  (Eqn.'s \ref{eq:dipole_responses}).}
         \label{fig:mfm_signal_dipole}
\end{figure}

A magnetic force microscope \cite{martin1987mif,rugar1990mfm,hartmann1999mfm} images small changes in the resonance frequency of a cantilever due to the interaction between magnetic material on a sharp tip at the end of the cantilever and local sample magnetic fields. The signal from a magnetic force microscope is proportional to $d F_{tip,z}/d z$, the derivative of the $z$-component of the force on the tip with respect to its $z$-position. Since this force sums contributions from all the magnetic material in the tip, the sensitivity and spatial resolution of MFM depends on the shape of the tip. For some applications and tip materials the tip can be modeled as having a magnetic monopole at its apex, with the force approximated by $F_{tip,z}=q B_z$, where $q$ is the effective magnetic $monopole$ moment of the tip \cite{schonenberger1990umf,bending1999lmp}. However, Lohau {\it et al.} have shown that their experimental MFM data can be modeled either using a monopole or a dipole approximation for the end of the tip, but that in either case the effective position of the monopole or dipole is a function of the field gradient from the sample \cite{lohau1999qde}. 

It is instructive to model the ultimate spatial resolution and sensitivity of an MFM in the point monopole and dipole approximations for the tip. The $z$-component of the force derivative on the tip from a field source can be written as \cite{lohau1999qde}
\begin{equation}
\frac{\partial F_z}{\partial z}= -\frac{q}{\mu_0} \frac{\partial B_z}{dz} + m_{t,x} \frac{\partial^2B_x}{\partial z^2}+m_{t,y}\frac{\partial^2B_y}{\partial z^2}+m_{t,z}\frac{\partial^2B_z}{\partial z^2},
\label{eq:lohau}
\end{equation}
where $q$ is the magnetic monopole flux, and $m_{t,i}$ are the dipole moments of the tip in the $x,y,z$ directions. Assume for simplicity that the tip is magnetized only in the $z$-direction. Then $m_{t,x}=m_{t,y}=0$ and the force derivatives due to the dipole, monopole, and current line sources of Eq.'s \ref{eq:dipole_field}, \ref{eq:monopole_field}, and \ref{eq:current_field} are given by
\begin{eqnarray}
\frac{\partial F_z}{\partial z} &=& \frac{3m_{t,z}\mu_0}{4\pi r^9} \left [ \right .   -5(m_{s,x}x+m_{s,y}y) z(3r^2-7z^2)  \nonumber \\
                                            &+& \left . m_{s,z}(3r^4-30r^2z^2+35z^4) \right ] \hspace{0.5in} {\rm Dipole-dipole} \nonumber \\
                                            &=& \frac{3\Phi_0m_{t,z}z}{2\pi } \frac{5z^2-3r^2}{r^7} \hspace{1.1in} {\rm Dipole-Monopole} \nonumber \\
                                            &=& -\frac{m_{t,z}\mu_0Ix}{\pi} \frac{x^2-3z^2}{(x^2+z^2)^3} \hspace{0.9in} {\rm Dipole-Current\,line}          
\label{eq:dipole_responses}
\end{eqnarray}   
where $m_{t,z}$ is the dipole moment of the tip.  

Similarly, the expressions for the force derivative in the monopole approximation are given by:  
\begin{eqnarray}
\frac{\partial F_z}{\partial z} &=& -\frac{3q}{4\pi r^7} \left [ \right .   (xm_{s,x}+ym_{s,y})(r^2-5z^2)  \nonumber \\
                                            &+& \left . m_{s,z}z(3r^2-5z^2) \right ] \hspace{1.22in} {\rm Monopole-dipole} \nonumber \\
                                            &=& -\frac{q\Phi_0}{2\pi \mu_0} \frac{r^2-3z^2}{r^5} \hspace{1.35in} {\rm Monopole-Monopole} \nonumber \\
                                            &=& \frac{qI}{\pi} \frac{xz}{(x^2+z^2)^2} \hspace{1.62in} {\rm Monopole-Current\,line}          
\label{eq:monopole_responses}
\end{eqnarray}   
                                                                                                  
The force derivative relations given by Eq.'s \ref{eq:dipole_responses} and \ref{eq:monopole_responses} are plotted in Figures \ref{fig:mfm_signal_dipole} and \ref{fig:mfm_signal_monopole} respectively.

\begin{figure}[tb]
        \begin{center}
                \includegraphics[width=16cm]{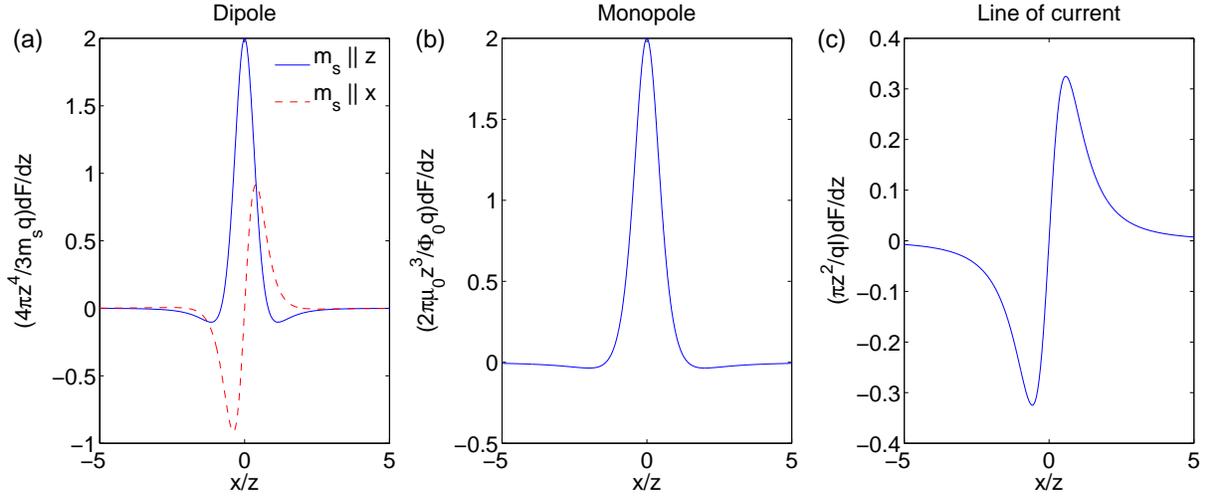}
        \end{center}
        \caption{MFM modeling using the monopole tip approximation: MFM force derivative curves for (a) a dipole source located at the origin with magnetic moment $m_s$ oriented parallel to $z$ (solid line) or parallel to $x$ (dashed line), Eq. \ref{eq:dipole_field}, (b) a monopole source located at the origin with magnetic flux $\Phi_0$, Eq. \ref{eq:monopole_field}, and (c) a line of current $I$ along the $y$ axis, Eq. \ref{eq:current_field}. The tip is modeled as a point monopole with magnetic flux $q$ scanning along the $x$ axis a height $z$ above the $xy$ plane (Eqn.s \ref{eq:monopole_responses}).} 
         \label{fig:mfm_signal_monopole}
\end{figure}

Within these approximations the force derivative signal strength increases strongly with decreasing $z$ - like 1/$z^5$ for a dipole source, 1/$z^4$ for a monopole source, and 1/$z^3$ for a current line source, in the point dipole tip approximation. In the monopole tip approximation the powers are 1/$z^4$, 1/$z^3$, and 1/$z^2$ for the three sources, respectively. The widths of the predicted force derivative features are proportional to the height $z$ of the tip above the source. The constant of proportionality does not depend strongly on source: the full width at half-maximum (or distance between the maximum and minimum force derivative for a line of current source) is $\Delta x/z$=0.6, 0.66, and 0.74 for the dipole tip, and 0.74, 1, and 1.16 for the monopole tip, for the three sources respectively. Of course, these conclusions would be modified for a real tip with a finite size. It has been estimated that the effective tip-sample spacing and radius of curvature for conventional MFM tips are both about 10 nm \cite{schonenberger1990umf}.

In order to estimate the sensitivity of MFM, one needs to know the minimum detectable force derivative, which depends on a number of characteristics of the MFM, including those of the cantilever. Using the lumped mass model for an MFM cantilever, the equation of motion of the tip end position $z$ is that of a driven, damped harmonic oscillator
\begin{equation}
m \frac{d^2 z}{dt^2}+\Gamma \frac{dz}{dt}+kz=F_{\rm signal}(t)+F_{\rm thermal}(t),
\label{eq:motion}
\end{equation}
where $m$ is the effective mass of the cantilever, $\Gamma=k/\omega_0Q$ is the damping constant, $k$ is the cantilever spring constant, $\omega_0$ is the angular resonant frequency, $Q$ is the quality factor, $F_{\rm signal}$ is the signal force (in this case the magnetic interaction between the tip and sample), and $F_{\rm thermal}$ is the force due to thermal fluctuations. The equipartition theorem implies that the power spectral density of thermal fluctuations $S_F=4\Gamma k_B T$. Then the thermally limited minimum detectable force derivative is given by
\begin{equation}
\frac{\partial F}{\partial z} _{\rm min} =\frac{1}{A} \sqrt{\frac{4 k k_B T BW}{\omega_0 Q}},
\label{eq:dfdzmin}
\end{equation}
where $BW$ is the measurement bandwidth and $A$ is the amplitude of oscillation of the cantilever.

\begin{figure}[tb]
        \begin{center}
                \includegraphics[width=10cm]{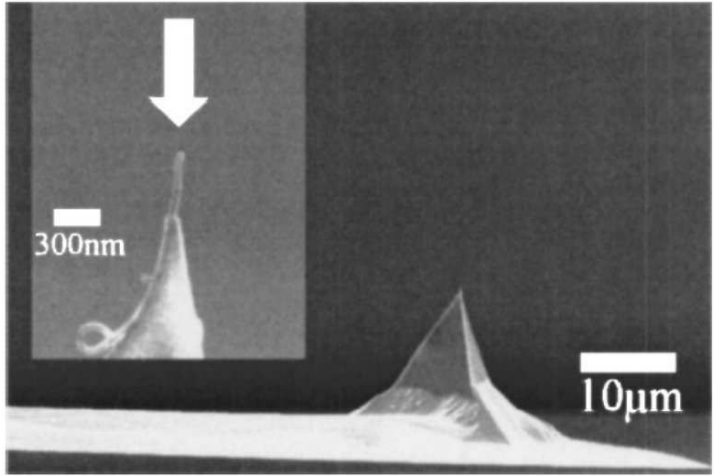}
        \end{center}
        \caption{Scanning electron microscope (SEM) image of a MFM cantilever. Insert upper left: high-resolution SEM image of the apex of the pyramid, where a coated carbon nanotube tip (CCNT) is visible. The arrow shows the direction of the metal evaporation. Reprinted figure with permission from Z. Deng, E. Yenilmez, J. Leu, J.E. Hoffman, E.W.J. Straver, H. Dai, and K.A. Moler,  \href{http://apl.aip.org/applab/v85/i25/p6263_s1}{Appl. Phys. Lett. {\bf 85}, 6263 (2004)}. Copyright 2004 by the American Institute of Physics.} 
         \label{fig:deng}
\end{figure}

A major thrust in research in MFM is to reduce the spatial extent of the magnetic material in the cantilever tip to improve spatial resolution and spin sensitivity. This is often done by shaping previously deposited material using focussed ion beam etching \cite{phillips2002hrm,litvinov2002osm}. Sharply defined magnetic tips can also be produced by evaporating magnetic material on an electron beam deposited carbon and ion etched needle \cite{koblischka2003ilr}, or onto a carbon nanotube \cite{deng2004mcc} at the end of a conventional Si cantilever tip (see Figure \ref{fig:deng}). Small amounts of magnetic material can also be deposited directly onto cantilever tips through nanoscale holes fabricated in a stencil mask \cite{champagne2003nss}. One approach would be to place nano-magnets directly on the apex of cantilever tips. This would serve three goals: improved spatial resolution, improved spin sensitivity, and improved interpretability,  since the magnetic moment of the deposited nano-magnets would have a narrow distribution \cite{sun2000mfn}. However, such tips could be especially susceptible to switching of the tip moment in sufficiently strong magnetic fields, reversing the contrast of the MFM images \cite{kirtley2007msn}. Although this complicates the interpretation of the images, the tip magnetization state is also known better, since the tip is generally close to saturation either up or down except for a narrow window around the switching points. The tip in MFM exerts a relatively strong magnetic field, comparable to the saturation magnetization of the magnetic material used in the tip, on the sample. This can cause switching of the local magnetic moment of the sample \cite{abraham1990tmf}, but can also be taken advantage of for vortex manipulation studies (see Section \ref{sec:manipulation}). 

Assume for the moment that it will be possible to locate small nanoparticles at the end of an MFM cantilever tip. A 7 nm diameter cobalt nanoparticle will have a total dipole moment of $m_t = 2.5 \times 10^{-19}A-m^2$, assuming a magnetization of $M=1.4\times 10^{6}$ A/m \cite{otani2000sii}. Taking values of $Q$=50,000, $\omega_0$=2$\pi \times$50 kHz, $A$=1 nm, $k$=2N/m, and  $T$=4K, the thermally limited force gradient is $\partial F/\partial z|_{\rm min}$=1.7$\times 10^{-7}$N/m. Using the dipole approximation for the tip, the peak force due to a sample dipole $m_s$ with both tip and sample moments oriented parallel to the $z$ axis is given by (Eq. \ref{eq:dipole_responses}, Fig. \ref{fig:mfm_signal_dipole}) $\partial F/\partial z_{peak}=24\mu_0m_sm_t/4\pi z^5$. Using $z$=15 nm results in a minimum detectible sample moment of 23 Bohr magnetons. From Fig. \ref{fig:mfm_signal_dipole} the full width at half maximum of the force derivative response would be $\Delta x = 0.6 z = 9$nm. 

\subsection{Summary}

Each of the three techniques described here has advantages and disadvantages: MFM has the best spatial resolution ($\sim$10-100 nm), SSM the worst ($\sim$ 0.3-10$\mu m$), with Hall bars intermediate ($\sim$ 0.1-5$\mu m$). The relative sensitivities of the three techniques depend on the type of field source. As in all microscopies there are tradeoffs between sensitivity and spatial resolution. The spatial resolution for SQUIDs and Hall bars is roughly given by $\Delta x \approx d$ for heights $z<<d$, where $d$ is the sensor size (Fig.'s \ref{fig:threesrc}, \ref{fig:squid_flux_and_width_vs_z}). For MFM the spatial resolution is roughly set by the tip-sample spacing $z$, in the limit where the radius of curvature of the tip end is smaller than $z$ (Fig.'s \ref{fig:mfm_signal_monopole},\ref{fig:mfm_signal_dipole}). The flux sensitivity of SQUIDs is nearly independent of SQUID or pickup loop size, so the field sensitivity is roughly proportional to the inverse of the pickup loop or SQUID area. With Hall bars limited by thermal noise the field sensitivity is independent of sensor size, but Hall bars dominated by $1/f$ noise are expected to have a field sensitivity that scales like $1/d$. Current MFM's measured field gradients, rather than fields, and the tradeoff between sensitivity and spatial resolution is less clear than for SSM and SHM, although larger amounts of magnetic material will  produce larger sensitivity but will require a larger tip volume, reducing resolution. Table \ref{tab:comparison} compares sensitivity and spatial resolution for a state of the art SQUID, with a pickup loop 0.6$\mu$m in diameter and a flux noise of 0.7$\times 10^{-6}\Phi_0/Hz^{1/2}$ (Figure \ref{fig:koshnick_squid} \cite{koshnick2008ats}), a state of the art Hall bar with sensor size 100 nm and a field noise of 500$\mu$T/$Hz^{1/2}$ at 3 Hz (Fig. \ref{fig:hicks_hallbar} \cite{hicks2007ncn}, and a hypothetical MFM with a 7 nm Co nanoparticle at the tip as discussed in Section \ref{sec:mfm_fundamentals}. These numbers should be treated with a great deal of caution. For example, I have assumed that $z/d=0.1$ for the SQUID and SHM, and $z$=15 nm for the MFM. These may be unreasonably optimistic estimates. However, the Table \ref{tab:comparison} does indicate how relative sensitivities depend on the field source. As an example, SSM and MFM have roughly comparable spin sensitivities, but SSM has nearly two orders of magnitude better sensitivity for a monopole source, and nearly 4 orders of magnitude better sensitivity for a line of current source. 

\begin{table}[h]
\caption{Sensitivities and spatial resolutions} %title of the table
\centering % centering table
\begin{tabular}{c ccccccc} % creating seven columns
\hline\hline %inserting double-line
&  &\multicolumn{2}{c}{Dipole} & \multicolumn{2}{c} {Monopole} &  \multicolumn{2}{c}{Line of current} \\ [0.5ex]
& & $N$($\mu_B/\sqrt{Hz}$)&$\Delta x$(nm)&$\Phi$($\Phi_0/\sqrt{Hz}$)&$\Delta x$(nm)&I(A/$\sqrt{Hz}$)&$\Delta x$(nm) \\ [0.5ex]
\hline \\ % inserts single-line
MFM && 23 & 9 & 3.4$\times10^{-5}$ & 10  & 2.6$\times10^{-5}$ & 11 \\ [0.5ex]
SHM && 4.5$\times 10^4$ & 100 & 3$\times 10^{-3}$ & 108 & 1.14$\times 10^{-4}$ & 118 \\ [0.5ex]
SSM && 78 & 600 & 8.8$\times10^{-7}$ & 650 & 5.5$\times10^{-9}$ & 710 \\ [0.5ex]
\hline % inserts single-line
\end{tabular}
\label{tab:comparison}
\end{table}

Table \ref{tab:scaling} provides a summary of the scaling exponents $n$ for the minimum detectable field sources for SSM, SHM, and MFM. In this table the minimum detectable dipole moment, flux, and current are proportional to $d^n$ for SSM and SHM, while for MFM these quantities are proportional to $z^n$. For SSM and SHM it is assumed that $z/d$ is held constant. For SSM it is assumed that thermal noise dominates. There are two columns for SHM, for noise dominated by thermal fluctuations, or by $1/f$ noise. For MFM there are two columns, for the monopole and the dipole tip approximations.  

\begin{table}[h]
\caption{Scaling exponents n} %title of the table
\centering % centering table
\begin{tabular}{c cccccc} % creating seven columns
\hline\hline %inserting double-line
 & SSM & Hall & Hall & MFM & MFM  \\ [0.5ex]
 &          & 1/f   & thermal & monopole & dipole \\
\hline \\ % inserts single-line
Line of current & -1 & 0 & 1 & 2 & 3   \\
Monopole        &  0 & 1  & 2 & 3 & 4   \\
Dipole             &   1 & 2  & 3 &  4 & 5  \\ 

\hline % inserts single-line
\end{tabular}
\label{tab:scaling}
\end{table}

There are further considerations in deciding which magnetic microscopy is best for a given application: SSM requires a cooled sensor, about 9K for the most technically advanced Nb SQUIDs, although Hall bars and MFM also have the highest sensitivities at low temperatures. SSM is the most straightforward to calibrate absolutely. Hall bars tend to be mechanically delicate and the most sensitive to destruction by electrostatic discharge.  MFM applies the largest fields to the sample. In general Hall bars and MFM can tolerate larger applied magnetic fields than SQUIDs. As we shall see in Section \ref{sec:applications}, each has been very useful in the study of superconductors.

\section{Applications}
\label{sec:applications}

\subsection{Static imaging of magnetic flux}

\begin{figure}[tb]
        \begin{center}
                \includegraphics[width=10cm]{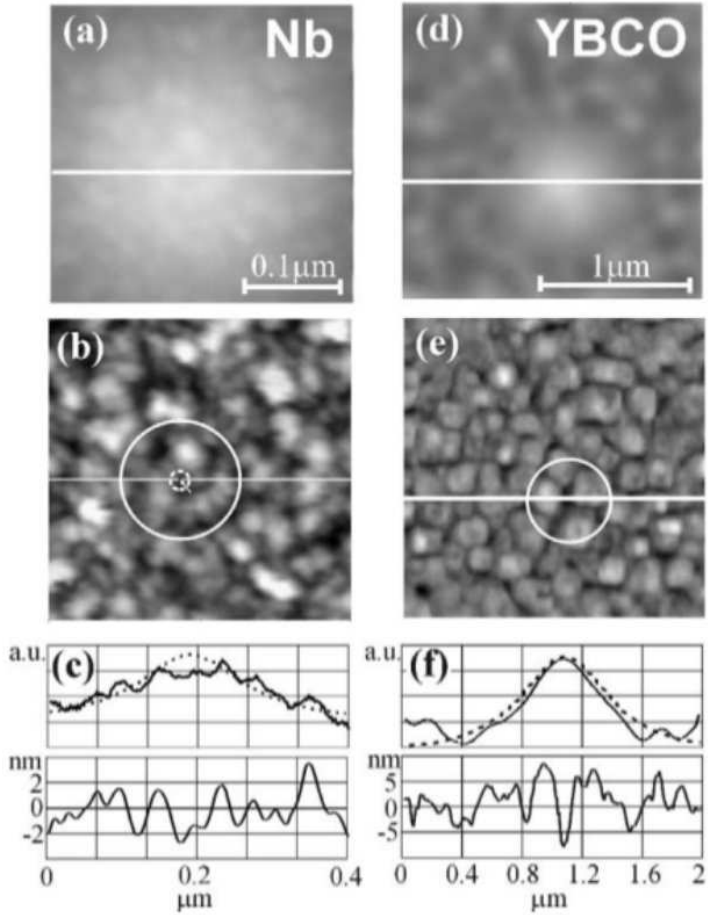}
        \end{center}
        \caption{(a) Magnetic force microscope image of 0.3 $\mu$m $\times$ 0.3 $\mu$m areas for a Nb film and the corresponding topographic image (b). (d) MFM image of a single vortex (2 $\times$ 2 $\mu m^2$) in a YBCO film and (e) the corresponding topographic image. The white circles in (b) and (e) correspond to the distance at which the stray field emanating from the vortex has decreased to 1/$e$ of its maximum value. The small dotted circle in (b) defines an area with diameter 2$\xi \approx$ 22 nm. The lower panels (c) and (f) are cross-sections of the MFM data and the topography profiles along the white lines indicated in (a), (b), (d), and (e). The dotted curves correspond to a Gaussian profile fitting of the MFM signal after filtering with a low-pass filter. Reprinted figure from \href{http://www.elsevier.com/}{Physica C: Superconductivity} {\bf 369}, A. Volodin, K. Temst, Y. Bruynseraede, C. van Haesendonck, M.I. Montero, I.K. Schuller, B. Dam, J.M. Huijbregtse and R. Griessen, ``Magnetic force microscopy of vortex pinning at grain boundaries in superconducting thin films", P. 165-170 (2002), with permission from Elsevier. } 
         \label{fig:volodin}
\end{figure}

Understanding the trapping of vortices in superconductors is of crucial technological importance, both because vortex pinning is the primary mechanism for enhanced critical currents in type II superconductors \cite{blatter1994vht}, and also because trapped vortices are an important source of noise in superconducting electronic devices. Superconducting vortices are especially easy to image using magnetic imaging because they are highly localized, quantized, and have relatively large magnetic fields. A number of different techniques have been used for imaging superconducting vortices, including SSM, SHM, and MFM. MFM has the advantage for this application that it can image both the vortex magnetic fields and sample topography simultaneously, so that macroscopic pinning sites can be identified. It has been used for a number of years to image superconducting vortices  \cite{hug1994omv,volodin2002mfm,moser1995osv,yuan1996vit,moser1998ltm,volodin2000ivc,volodin2000mfm2,roseman2002mia}. 
Care must be taken however, because the large magnetic fields exerted by the tip can distort and dislodge vortices. As we will see in Section \ref{sec:manipulation}, this can also be an advantage.  An particularly striking example of simultaneous magnetic and topographic imaging of vortices in Nb and YBCO is shown in Figure \ref{fig:volodin} \cite{volodin2002mfm}. 
%Roseman {\it et. al} \cite{roseman2002dtv} used temperature dependent magnetic force microscopy to determine the critical temperature of a Nb film. 

\subsubsection{Narrow strips}

\begin{figure}[tb]
        \begin{center}
                \includegraphics[width=14cm]{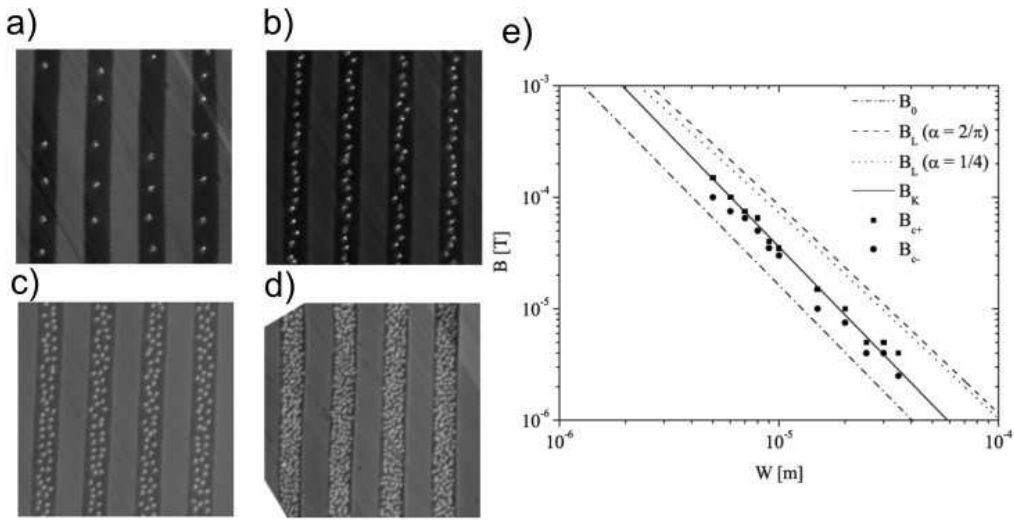}
        \end{center}
        \caption{SQUID microscope images of 35 $\mu$m wide YBCO thin film strips cooled in magnetic inductions of (a) 5, (b) 10, (c) 20, and (d) 50 $\mu$T.
        e) Critical inductions for vortex trapping in YBCO thin films as a function of strip width. The squares (B$_{c+}$) represent the lowest induction in which vortices were observed, and the dots (B$_{c-}$) are the highest inductions for which vortex trapping was not observed. The dashed-dotted line is the metastable critical induction B$_0$ [Eq. \ref{eq:clem}] \cite{clem1998ves}, the short-dashed and long-dashed lines are B$_L$ [Eq. \ref{eq:likharev}], the critical induction calculated using an absolute stability criterion, and the solid line is B$_K$ [Eq. \ref{eq:kuit}], calculated using a dynamic equilibrium criterion between thermally activation and escape for vortices. Reprinted figure with permission from K.H. Kuit, J.R. Kirtley, W. van der Veur, C.G. Molenaar, F.J.G. Roesthuis, A.G.P. Troeman, J.R. Clem, H. Hilgenkamp, H. Rogalla, and J. Flokstra,  \href{http://prb.aps.org/abstract/PRB/v77/i13/e134504}{Phys. Rev. B {\bf 77}, 134504 (2008)}. Copyright 2008 by the American Physical Society.}
        \label{fig:trapprb}
\end{figure}

The sensitivity of high-T$_c$ superconducting sensors such as SQUIDs and hybrid magnetometers based on high-T$_c$ flux concentrators is limited by $1/f$ noise. One source of this noise is the movement of vortices trapped in the sensor. One technique for reducing this source of noise is to trap vortices at a distance from the sensitive regions of the superconducting circuitry using holes or motes \cite{ketchen1985jcs,bermon1983mgj}. Jeffery {\it et al.} \cite{jeffery1995mim} used a scanning SQUID microscope to image flux trapping in superconducting electronic circuitry with motes. Veauvy {\it et al.} \cite{veauvy2004msm} studied vortex trapping in regular arrays of nanoholes in superconducting aluminum. Noise from trapped vortices in high-T$_c$ SQUID washers and flux concentrators can be eliminated by dividing the high-T$_c$ body into thin strips \cite{dantsker1997hts,jansman1999sht}. Below a critical magnetic induction no vortex trapping occurs in these strips for a given strip width. A number of models for the critical inductions of thin-film strips have been proposed  \cite{likharev1972,clem1998ves,kuznetsov1999ofp,maksimova1998msc}. Indirect experimental testing of these models was done by observing noise in high-T$_c$ SQUIDs as a function of strip width and induction \cite{dantsker1997hts,jansman1999sht,dantsker1996roo}. More direct experimental verification of these models was presented by Stan {\it et al.} \cite{stan2004cfc} using scanning Hall probe microscopy on Nb strips and Suzuki {\it et al.} \cite{suzuki2000min} using SQUID microscopy on NdBa$_2$Cu$_3$O$_y$ thin-film patterns with slots. Both experiment and theory found that the critical magnetic induction varied roughly as $1/W^2$, where $W$ is the strip width.  However, the experimental \cite{stan2004cfc} and theoretical \cite{clem1998ves,likharev1972} pre-factors multiplying this $1/W^2$ dependence differ significantly. 

Figure \ref{fig:trapprb}a-d shows SQUID microscope images of 35$\mu$m wide YBCO strips cooled in various magnetic inductions. Fig. \ref{fig:trapprb}e plots B$_{c+}$, the lowest induction at which vortices are trapped in the strips (squares), and B$_{c-}$, the highest induction at which vortices are not trapped (dots) as a function of strip width. The lines represent various theoretical predictions.

The Gibbs free energy for a vortex in a superconducting strip of width $W$ in an applied induction B$_a$ perpendicular to the strip plane can be written as \cite{kuit2008vte}:
\begin{equation}
G(x) = \frac{\Phi_0^2}{2\pi\mu_0\Lambda} \ln \left [ \frac{\alpha W}{\xi} \sin \left( \frac{\pi x}{W} \right ) \right ] \mp \frac{\Phi_0 (B_a-n\Phi_0)}{\mu_0\Lambda} x(W-x),
\label{eq:gibbs}
\end{equation}
where $\Lambda=2\lambda^2/d$ is the Pearl length of the strip, $\lambda$ is the London penetration depth, $\alpha$ is a constant of order 1, $n$ is the areal density of vortices, $\xi$ is the coherence length, and $x$ is the lateral position of the vortex in the strip. The critical induction model of Likharev \cite{likharev1972} states that in order to trap a vortex in a strip the vortex should be absolutely stable: the critical induction then happens when the Gibbs free energy in the middle of the strip is equal to zero, and leads to
\begin{equation}
B_{L}=\frac{2\Phi_0}{\pi W^2} \ln \left ( \frac{\alpha W}{\xi} \right ).
\label{eq:likharev}
\end{equation}
This is plotted as the dashed and dotted lines in Fig. \ref{fig:trapprb} for two assumed values of $\alpha$. A second model, proposed by Clem \cite{clem1998ves} considers a metastable condition, in which the induction is just large enough to cause a minimum in the Gibbs energy at the center of the strip, $d^2G(W/2)/dx^2=0$, leading to
\begin{equation}
B_0=\frac{\pi \Phi_0}{4 W^2}.
\label{eq:clem}
\end{equation}
This is plotted as the dot-dashed line in Fig. \ref{fig:trapprb}. Finally, Kuit {\it et al.} proposed that the critical induction should result from a dynamic equilibrium between vortex thermal generation and escape. This leads to the critical induction
\begin{equation}
B_K = 1.65 \frac{\Phi_0}{W^2}.
\label{eq:kuit}
\end{equation}
It can be seen in Figure \ref{fig:trapprb} that the dynamic equilibrium model (solid line) fits the data the best of the three models, with no fitting parameters. Kuit {\it et al.} also found that $n$, the areal density of trapped vortices, increased linearly with applied induction above the critical value, in agreement with the dynamic equilibrium model, and that the trapping positions showed lateral ordering, with a critical field for formation of a second row that was consistent with a numerical prediction of Bronson {\it et al.} \cite{bronson2006ecp}. Similar good agreement between experiment and the dynamic equilibrium model was found by the same authors for Nb thin film strips.

\subsubsection{Superconducting wire lattices, clusters,  and nano-hole arrays}

\begin{figure}[tb]
        \begin{center}
                \includegraphics[width=16cm]{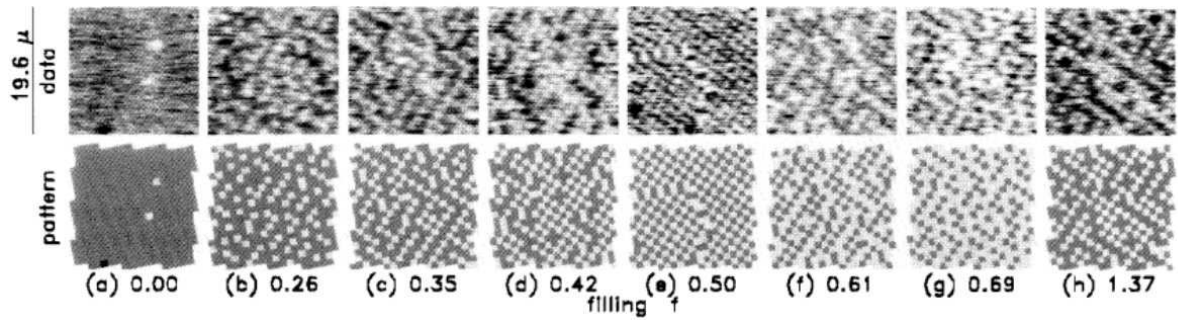}
        \end{center}
        \caption{Top row: Scanning Hall probe microscopy images of  vortex configurations in square areas 19.6 $\mu$m on a side in a superconducting Nb wire grid of 0.95 $\mu$m $\times$ 0.95 $\mu$m square holes, at the filling fractions $f$ indicated. The second row shows maps of the positions in the grid occupied by vortices. Note the ordering observed near $f=1/3$ (c), $f=1/2$ (e), and $f=4/3$ (h). Reprinted figure with permission from H.D. Hallen, R. Seshadri, A.M. Chang, R.E. Miller, L.N. Pfeiffer, K.W. West, C.A. Murray, and H.F. Hess,  \href{http://prl.aps.org/abstract/PRL/v71/i18/p3007_1}{Phys. Rev. Lett. {\bf 71}, 3007 (1993)}. Copyright 1993 by the American Physical Society.}
        \label{fig:hess}
\end{figure}

Two dimensional arrays of Josephson junctions \cite{webb1983mfb,voss1982pcw,sanchez1981pnn,resnick1981ktt,abraham1982rtt,fazio2001qpt} have been studied extensively, not only because they offer the possibility of studying Kosterlitz-Thouless effects \cite{kosterlitz1973omp} in an ordered 2-dimensional system, but also because of interesting effects that were predicted \cite{teitel1983jja} and observed \cite{webb1983mfb,voss1982pcw} to occur when the applied field is a rational fraction of the field required to populate each cell with a quantum of magnetic flux. 
Scanning magnetic microscopy has been used extensively for studying vortex trapping in two-dimensional superconducting wire networks \cite{hallen1993dsi,vu1993imv}, clusters \cite{vu1993imv}, and ring arrays \cite{davidovic1996cda,davidovic1997mcg}.  Superconducting wire networks are of interest because they are a physical realization of a frustrated $xy$ model \cite{teitel1983ptf,teitel1983jja,kolahchi1991gsu}, where the variable is the phase of the superconducting order parameter. The filling fraction $f$ of vortices/site can be varied by cooling the network in various fields, and detailed predictions have been made for the ground state vortex configurations at rational $f$ fractions \cite{halsey1985jja}. Studies of these systems were at first performed using transport \cite{pannetier1984eft,itzler1990csd,rzchowski1990vpj,lobb1983tir}, but direct imaging of the vortex positions provides more detailed information. Runge {\it et al.} did magnetic decoration experiments on superconducting wire lattices \cite{runge1993fds}. Although these experiments provided direct information on the vortex positions, only one set of conditions, such as magnetic field and temperature, can be explored for each sample using this technique. An example of scanning Hall bar microscopy on a superconducting wire network is shown in Figure \ref{fig:hess} \cite{hallen1993dsi}. In this figure ordered vortex arrangements can be seen near the fractional filling factors of $f$=1/3, 1/2, and 4/3, as predicted by Teitel and Jayaprakash \cite{teitel1983ptf,teitel1983jja}, along with grain boundaries between ordered domains and vacancies.

Field {\it et al.} \cite{field2002vcm} studied vortex trapping in Nb films with large arrays of 0.3$\mu$m holes on a square lattice. They found strong matching effects when the cooling fields corresponded to one or two vortices per site, as well as at several fractional multiples of the matching field. They also observed striking domain structure and grain boundaries between the domains.

\subsubsection{Half-integer flux quantum effect}
\label{sec:symmetry}

The discovery in 1986 of superconductivity at high temperatures in the cuprate perovskites \cite{bednorz1986pht,wu1987san} generated enormous excitement. Although there was a great deal of evidence that the superconducting gap in the cuprates was highly anisotropic, with a significant density of states at low energies \cite{scalapino1995cdp,annett96},  phase sensitive measurements \cite{vanHarlingen1995pst,tsuei2000psc} were required to distinguish, for example, $d$-wave from highly anisotropic $s$-wave pairing symmetry. The first such experiments \cite{wollman1993eds,wollman1995edp} used the magnetic field dependence of the critical current in single junctions and 2-junction SQUIDs between single crystals of YBa$_2$Cu$_3$O$_{7-\delta}$ (YBCO) and Pb to demonstrate a $\pi$-phase shift between the component of the superconducting pairing order parameter perpendicular to adjacent in-plane crystal faces ({\it e.g.}, ignoring the effects of twinning, between the $a$ vs $b$ axis normal faces)  of the YBCO. Similar phase sensitive Josephson interference experiments were performed by Brawner {\it et al.} \cite{brawner1994eus}, Iguchi {\it et al.} \cite{iguchi1994eed}, Miller {\it et al.} \cite{miller1995utj} and Kouznetsov {\it et al.} \cite{kouznetsov1997caj}. 

\begin{figure}
        \begin{center}
                \includegraphics[width=10cm]{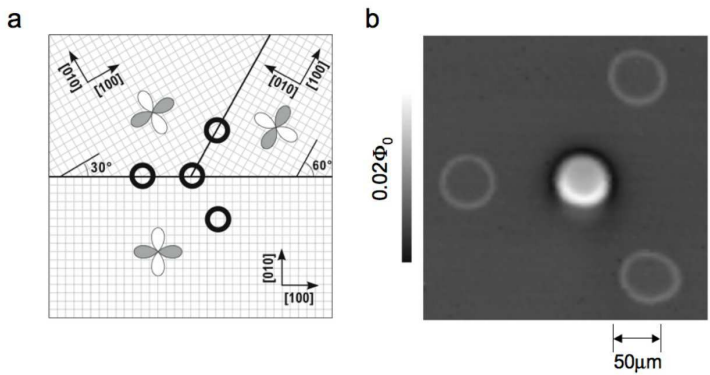}
        \end{center}
        \caption{(a) Schematic diagram for the tricrystal (100) SrTiO$_3$ substrate used in the phase sensitive pairing symmetry tests of Tsuei et al. \cite{tsuei1994psf}. Four epitaxial YBa$_2$Cu$_3$O$_{7-\delta}$ thin film rings are interrupted by 0, 2, or 3 grain boundary Josephson weak links. (b) Scanning SQUID microscopy image of the four superconducting rings in (a), cooled in a field $<$5 mG. The central ring has $\Phi_0/2$  magnetic flux, where $\Phi_0$ is the superconducting flux quantum, spontaneously generated in it. The other 3 rings, which have no spontaneously generated flux, are visible through a small change in the self-inductance of the SQUID when passing over the superconducting walls of the rings.}

        \label{fig:tricrystal}
\end{figure}

A  phase-sensitive technique for determining pairing symmetry that is complementary to Josephson interference is to image magnetic fields generated by spontaneous supercurrents in superconducting rings that have an odd number of intrinsic $\pi$-phase shifts in circling them, or in Josephson junctions that have intrinsic $\pi$-shifts along them \cite{ geshkenbein1986jes,geshkenbein1987vhm}. The first such experiments were done by Tseui {\it et al.} \cite{tsuei1994psf}, using YBCO films grown epitaxially on tricrystal SrTiO$_3$ substrates. A scanning SQUID microscope was used to image the magnetic fields generated by supercurrents circulating rings photolithographically patterned in the YBCO films. In the original experiments the central ring  (Fig. \ref{fig:tricrystal}), which has 3 grain boundary Josephson junctions, has either 1 or 3 intrinsic $\pi$-phase shifts for a superconductor with predominantly $d_{x^2-y^2}$ pairing symmetry, and should therefore have states of local energy minima, when no external magnetic field is applied, with the total flux through the ring of $\Phi=(n+1/2)\Phi_0$, n an integer, for sufficiently large $LI_c$ products, where $L$ is the inductance of the ring, and $I_c$ is the smallest critical current of the junctions in the ring. This is the half-integer flux quantum effect. The two outer rings that cross the tricrystal grain boundaries have 0 or 2 intrinsic $\pi$-phase shifts for a $d_{x^2-y^2}$ superconductor, and should therefore have local energy minima at $\Phi=n\Phi_0$ - conventional flux quantization. Similarly, the ring that does not cross any grain boundaries has no intrinsic phase drop, and also has local minimum energies for $\Phi=n\Phi_0$. These expectations were confirmed using a SQUID microscope using a number of techniques \cite{tsuei1994psf}, consistent with YBCO having $d_{x^2-y^2}$ pairing symmetry. Subsequent tricrystal experiments  eliminated the possibility of a non-symmetry related origin for the half-integer flux quantization in the 3-junction ring \cite{kirtley1995sop}, ruled out extended $s$-wave symmetry \cite{tsuei1995fqt}, demonstrated the existence of half-flux quantum Josephson vortices \cite{kirtley1996dii}, showed results consistent with $d_{x^2-y^2}$ pairing symmetry for the hole-doped high-T$_c$ cuprate superconductors Tl$_2$Ba$_2$CuO$_{6+\delta}$ \cite{tsuei1996hiq},  {Bi}$_2${Sr}$_2${CaCu}$_2${O}$_{8+\delta}$ \cite{kirtley1996hif}, and La$_{2-x}$Sr$_x$CuO$_{4-y}$ \cite{tsuei2004rdp} , for a range of doping concentrations \cite{tsuei2004rdp}, and the electron-doped cuprates    Nd$_{1.85}$Ce$_{0.15}$CuO$_{4-y}$  (NCCO) and Pr$_{1.85}$Ce$_{0.15}$CuO$_{4-y}$ (PCCO) \cite{tsuei2000pse}, and for a large range in temperatures \cite{kirtley1999tdh}.  SQUID magnetometry experiments on two-junction YBCO-Pb thin film SQUIDs were performed by Mathai {\it et al.} \cite{mathai1995ept}, and on tricrystals by Sugimoto {\it et al.} \cite{sugimoto2002tdh}. Josephson interference, SQUID magnetometry, and microwave measurements were made on biepitaxial all high-T$_ c$ YBCO junctions by Cedergren {\it et al.} \cite{cedergren2010ibs}.

\begin{figure}
        \begin{center}
                \includegraphics[width=10cm]{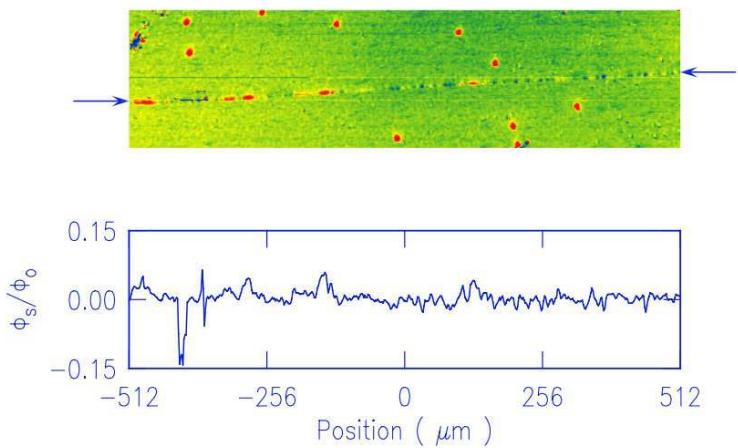}
        \end{center}
        \caption{Scanning SQUID microscope image of an area of 1024 $\times$ 256 $\mu$m$^2$ along an asymmetric 45$^o$ [001] tilt YBa$_2$Cu$_3$O$_{7-\delta}$ bicrystal grain boundary (arrows). There are some ten bulk vortices in the grains (film thickness $\sim$180 nm), but there is also flux of both signs spontaneously generated in the grain boundary. The bottom section shows a cross section through the data along the grain boundary measured in units of $\Phi_0$ penetrating the SQUID pickup loop. Reprinted figure with permission from J. Mannhart, H. Hilgenkamp, B. Mayer, Ch. Gerger, J.R. Kirtley, K.A. Moler, and M. Sigrist, \href{http://link.aps.org/abstract/PRL/v77/p2782}{Phys. Rev. Lett. {\bf 77}, 2782 (1996)}. Copyright 1996 by the American Physical Society.}

        \label{fig:mannprl3}
\end{figure}

Grain boundary junctions in the high-T$_c$ cuprate superconductors are Josephson weak links.  As such they are used for fundamental and device applications. In addition, grain boundaries play an important role in limiting the critical current density in superconducting cables made from the cuprates \cite{hilgenkamp1992gbh}. Anomalies in the dependence on magnetic field of the critical current of cuprate grain boundary junctions with misorientation angles near 45$^\circ$ can be understood in terms of their $d_{x^2-y^2}$ pairing symmetry combined with faceting, which naturally occurs on a 10-100nm scale in these grain boundaries \cite{hilgenkamp1996ids}. This combination produces a series of intrinsic $\pi$-phase shifts along such grain boundaries. These $\pi$-shifts have spontaneous supercurrents and magnetic fields associated with them \cite{mannhart1996gmf}. An example of SQUID microscope imaging of these spontaneous fields is shown in Figure \ref{fig:mannprl3}. If the distance between intrinsic $\pi$-shifts is shorter than the Josephson penetration depth, the total magnetic flux associated with each will be less than $\Phi_0/2$ \cite{kirtley1997sfm}; a high density of facets can result in ``splintered" Josephson vortices with fractions of $\Phi_0$ of flux\cite{mints2002osj}.

\begin{figure}[tb]
        \begin{center}
                \includegraphics[width=10cm]{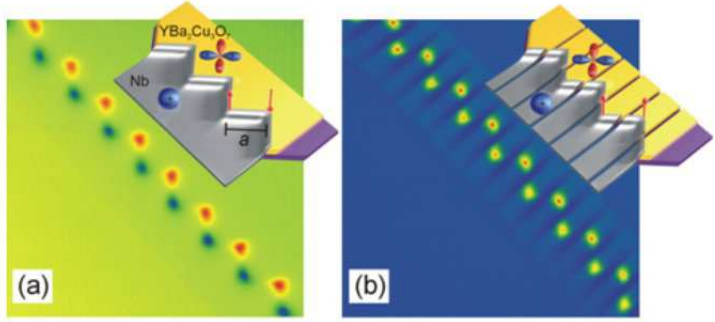}
        \end{center}
        \caption{Generation of half-flux quanta in connected and unconnected YBa$_2$Cu$_3$O$_{7-\delta}$-Au-Nb zigzag structures. Shown are scanning SQUID micrographs of (a) 16 antiferromagnetically ordered half-integer flux quanta at the corners of a connected zigzag structure, and (b) 16 ferromagnetically ordered half-flux quanta at the corners of an unconnected zigzag structure (T=4.2K).  The corner to corner distance is 40 $\mu$m in both (a) and (b). Reprinted figure with permission from Macmillan Publishers Ltd: H. Hilgenkamp, Ariando, H.J.H. Smilde, D.H.A. Blank, G. Rijnders, H. Rogalla, J.R. Kirtley, and C.C. Tsuei, \href{http://www.nature.com/nature/journal/v422/n6927/abs/nature01442.html}{Nature {\bf 422}, 50 (2003)}. Copyright 2003. }

        \label{fig:hilgnat1}
\end{figure}

Facetted (zigzag) junctions can also be fabricated by design, using for example a ramp-edge YBCO/Nb junction technology \cite{smilde2002etr}. An example is shown in Figure \ref{fig:hilgnat1}, which shows scanning SQUID microscope images of two facetted junctions \cite{hilgenkamp2003omm}. In Fig. \ref{fig:hilgnat1}a the corner junctions are connected by superconducting material. In this case there is a strong interaction between the half-flux quantum Josephson vortices upon cooling, so that they have a strong tendency to align with alternating signs of the circulating supercurrents and spontaneous magnetizations. In Fig. \ref{fig:hilgnat1}b the corner junctions have been isolated from each other by etching channels between them. In this case the half-flux quantum Josephson vortices all have the same alignment, even when cooled in a relatively small field.

\begin{figure}[tb]
        \begin{center}
                \includegraphics[width=10cm]{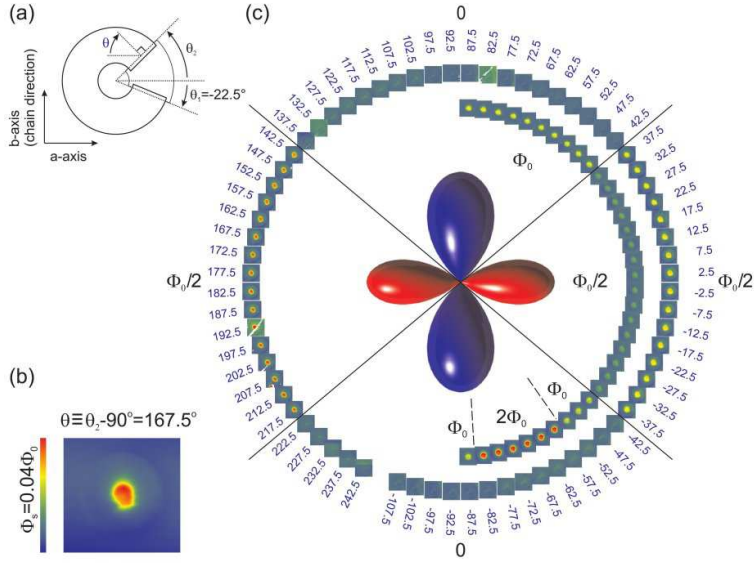}
        \end{center}
        \caption{(a) Schematic of YBCO-Nb two-junction rings fabricated to phase sensitively determine the momentum dependence of the in-plane gap in YBCO. (b) SQUID microscope image, taken with a square pickup loop 8 $\mu$m on a side, of a square area 150 $\mu$m on a side centered on one of the rings, cooled and imaged in zero field at 4.2K.  (c) The outer circle of images, taken after cooling in zero field, has a full-scale variation of 0.04 $\Phi_0$ flux through the SQUID: the inner ring, taken after the sample was cooled in a field of 0.2 $\mu$T, has a full scale variation of 0.09 $\Phi_0$. The rings cooled in zero field had either 0 or $\Phi_0/2$ of flux in them; the rings cooled in 0.2 $\mu$T had $\Phi_0/2$, $\Phi_0$ or 2$\Phi_0$ of flux, as labelled. Reprinted figure with permission from Macmillan Publishing Ltd: J.R. Kirtley, C.C. Tsuei, Ariando, C.J.M. Verwijs, S. Harkema and H. Hilgenkamp,  \href{http://www.nature.com/nphys/journal/v2/n3/abs/nphys215.html}{Nature Physics {\bf 2}, 190 (2006)}. Copyright 2006.}
        \label{fig:varma_fig_2}
\end{figure}

Lombardi {\it et al.} \cite{lombardi2002idw}, and Smilde {\it et al.} \cite{smilde2005adw} have mapped out the in-plane momentum dependence of the superconducting gap amplitude of YBCO by measuring the critical current as a function of junction normal angle for a series of biepitaxial YBCO grain boundary junctions, and YBCO/Nb ramp junctions respectively. The latter experiments, performed in part on untwinned epitaxial YBCO films, were able to determine the anisotropy of the gap between the $a$ and $b$ in-plane crystalline directions. However, these experiments were only sensitive to the amplitude of the superconducting order parameter, not its phase. Phase sensitive measurements of the momentum dependence of the in-plane gap in YBCO were performed by Kirtley {\it et al.} \cite{kirtley2006arp} by imaging the spontaneously generated magnetic flux in a series of YBCO-Nb two-junction rings, with one junction angle normal relative to the YBCO crystalline axes held fixed from ring to ring, while the other junction normal angle varied. Some results from this study are shown in Figure \ref{fig:varma_fig_2}.  The rings either had $n\Phi_0$ (integer flux quantization) or $(n+1/2)\Phi_0$ (half-integer flux quantization) of flux in them, $n$ an integer. The transition between integer and half-integer flux quantization happened at junction angles slightly different from 45$^\circ$+$n\times90^\circ$, as would be expected for a pure $d_{x^2-y^2}$ superconductor, because of the difference in gap amplitudes between the crystalline $a$ and $b$ phases. In addition, careful integration of the total flux in the rings showed that any imaginary component to the order parameter, if present, must be small.

\subsubsection{Ring arrays}

\label{sec:ring_arrays}

\begin{figure}[tb]
        \begin{center}
                \includegraphics[width=10cm]{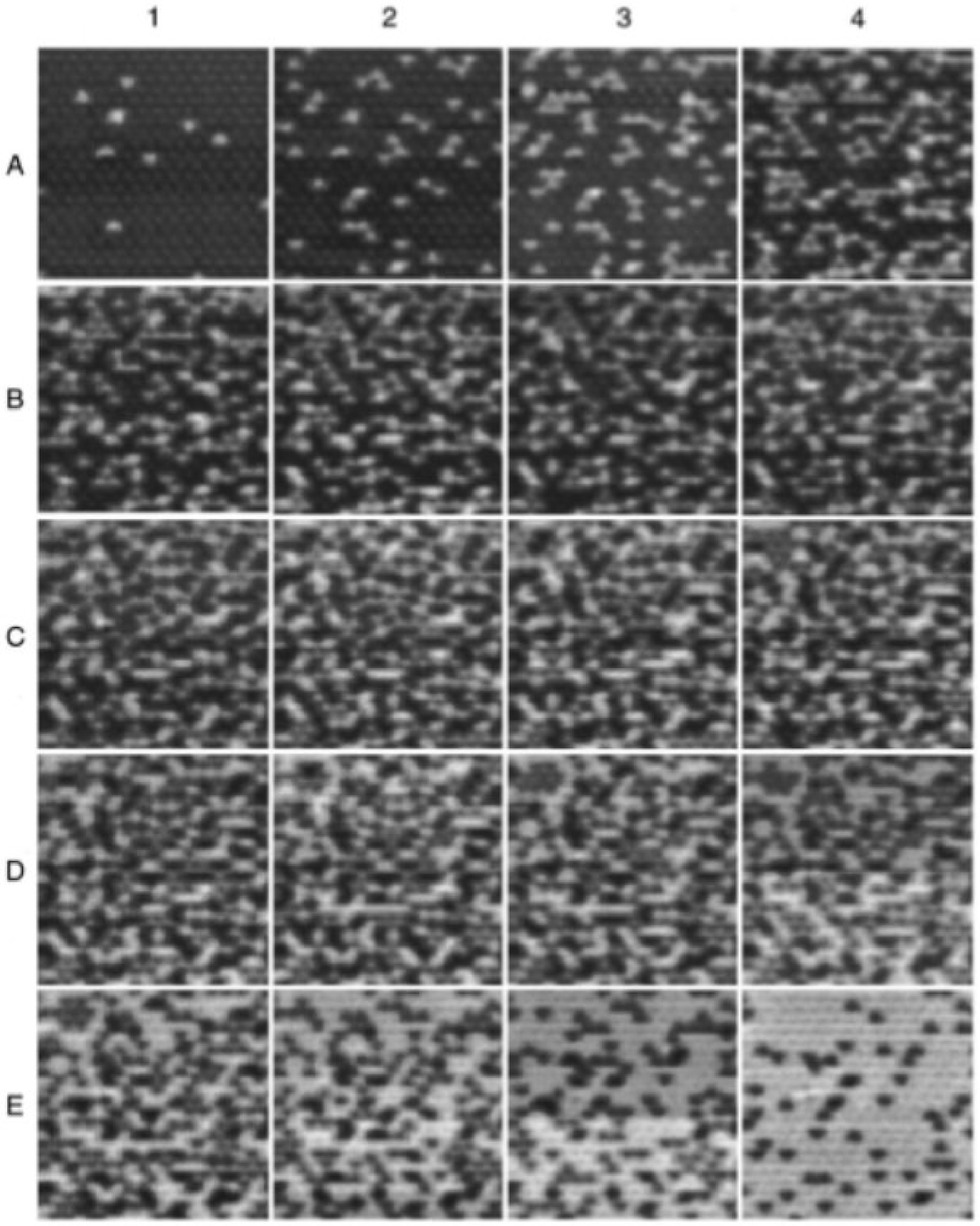}
        \end{center}
        \caption{A sequence of field-cooled images of a honeycomb lattice of 1$\mu$m diameter hexagonal Nb thin film rings, taken in increasing applied fluxes near an applied flux per ring of $\Phi_0/2$. The flux increases from left to right starting at the upper left corner with 0.4913$\Phi_0$ and ending in the lower right corner at 0.5066$\Phi_0$. Reprinted figure with permission from D. Davidovi{\'c}, S. Kumar, D.H. Reich, J. Siegel, S.B. Field, R.C. Tiberio, R. Hey, and K. Ploog,  \href{http://prb.aps.org/abstract/PRB/v55/i10/p6518_1}{Phys. Rev. B {\bf 55}, 6518 (1997)}. Copyright 1997 by the American Physical Society.} 
         \label{fig:davidovic_fig_14}
\end{figure}

\begin{figure}[tb]
        \begin{center}
                \includegraphics[width=10cm]{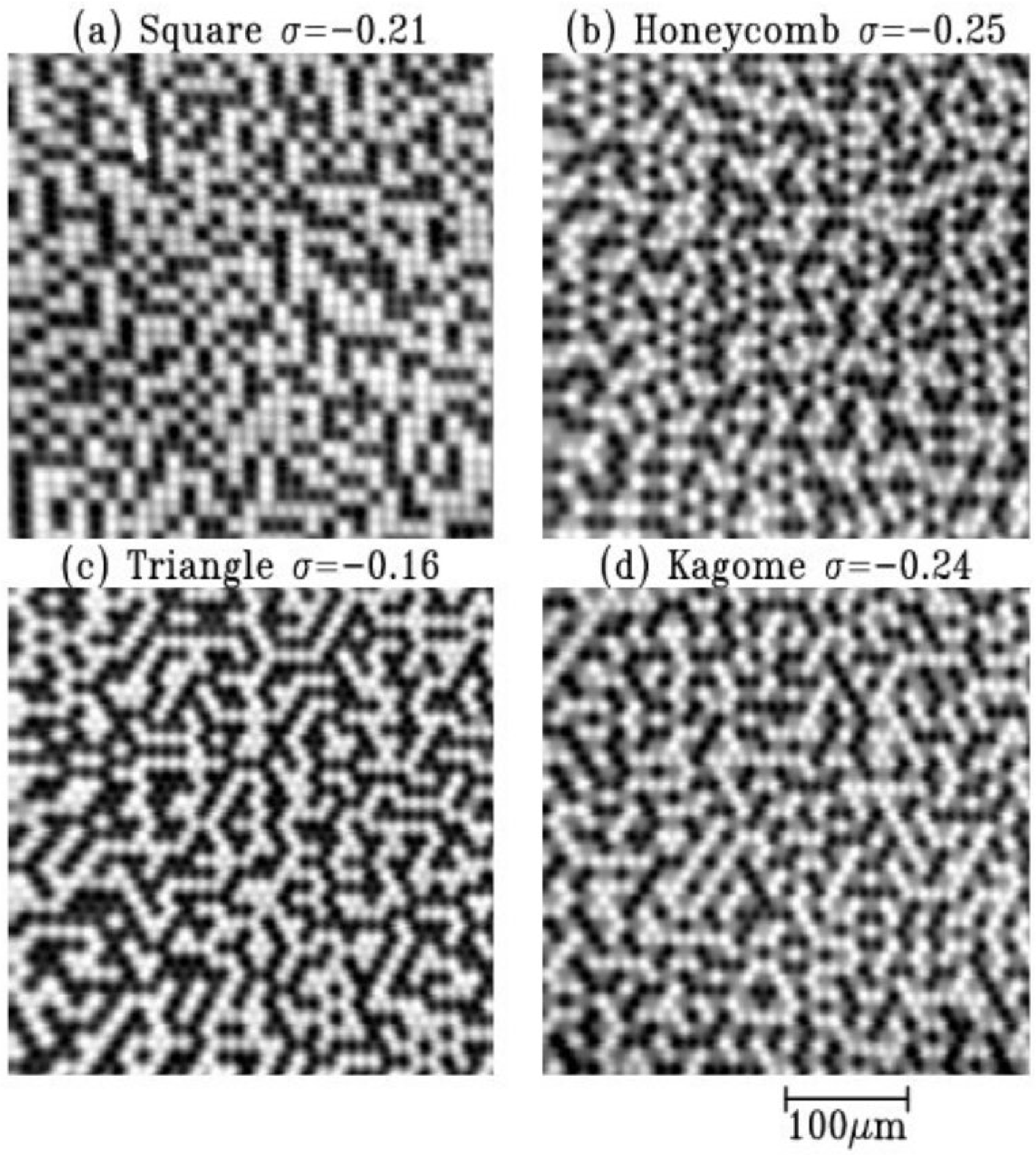}
        \end{center}
        \caption{SQUID microscopy images of four electrically disconnected arrays of $\pi$-rings with 11.5 $\mu$m nearest neighbor distances. These images were taken at 4.2 K with a 4 $\mu$m diameter pickup loop after cooling in nominally zero field at 1-10 mK/s. The bond-order $\sigma$ is a measure of the anti-ferromagnetic order in the arrays. The geometrically unfrustrated square and honeycomb arrays could have perfect antiferromagnetic correlation, which would correspond to $\sigma=-1$. The minimum possible bond-order for the frustrated Kagome and triangle lattices is $\sigma=-1/3$. Reprinted figure with permission from J.R. Kirtley, C.C. Tsuei, Ariando, H.J.H. Smilde, and H. Hilgenkamp,  \href{http://prb.aps.org/abstract/PRB/v72/i21/e214521}{Phys. Rev. B {\bf 72}, 214521 (2005)}. Copyright 2005 by the American Physical Society.}

        \label{fig:hilgprb6}
\end{figure}

Arrays of superconducting rings are of interest as a physical analogue of a spin: for conventional superconducting rings the states with shielding supercurrent flowing clockwise or counter-clockwise are degenerate at a magnetic flux bias of $\Phi_0/2$ (half of a flux quantum threading each ring). Two neighboring rings can interact magnetically, making them a physical analogue of the Ising spin. Frustration can be introduced into the system by choosing the arrangement of the rings: square and hexagonal lattices are geometrically unfrustrated, while triangular and Kagome lattices are frustrated. 

Davidovic {\it et al.} used scanning Hall bar microscopy to study two-dimensional arrays of closely spaced Nb rings (see Fig. \ref{fig:davidovic_fig_14}) \cite{davidovic1996cda,davidovic1997mcg}.  In these experiments Davidovic {\it et al.} found that it was possible to find small islands of antiferromagnetically ordered ``spins", but no perfect ordering, even in arrays with no geometrical frustration. One source of disorder in arrays of conventional superconducting rings is variation in area from ring to ring from lithographic variations. Conventional rings must be flux biased to near $\Phi_0/2$ flux with high precision (see Fig. \ref{fig:davidovic_fig_14}), and area variations lead to flux variations. 

The same technology used for the pairing symmetry tests of Fig. \ref{fig:varma_fig_2} were used to form two-dimensional arrays of $\pi$-rings, superconducting rings with an intrinsic phase shift of $\pi$ \cite{hilgenkamp2003omm}.  An example of SQUID microscope imaging of several $\pi$-ring arrays is shown in Figure \ref{fig:hilgprb6} \cite{kirtley2005aoa}. $\pi$-rings have two degenerate states at zero flux bias, therefore eliminating the need for precise flux biasing of the rings, and thereby eliminating slight differences in the sizes of the rings as a source of disorder \cite{kirtley2005aoa}. Although $\pi$-ring arrays showed more negative values of the bond-order $\sigma$, corresponding to greater antiferromagnetic ordering, than arrays of conventional rings, perfect ordering was never observed in either conventional or $\pi$-ring arrays, possibly because of the large number of nearly degenerate states in this system \cite{O'Hare2006esl}.

\subsubsection{Tests of the interlayer tunneling model}

One candidate mechanism for superconductivity at high temperatures in the cuprate perovskite superconductors is the interlayer tunneling model, in which the superconductivity results from an increased coupling between the layers in the superconducting state \cite{wheatley1988ieh,anderson1991coa,anderson1992ect,chakravarty1993itg,anderson1995itm}. An essential test of this theory is the strength of the interlayer Josephson tunneling in layered superconductors. For the interlayer tunneling model to succeed, the interlayer coupling in the superconducting state must be sufficiently strong to account for the large condensation energy of the cuprate superconductors. The best materials for testing this requirement are Tl$_2$Ba$_2$CuO$_{6+\delta}$ (Tl-2201) and HgBa$_2$CuO$_{4+\delta}$ (Hg-1201), which have high critical temperatures (T$_c \approx$90 K) and a single copper oxide plane per unit cell. Scanning SQUID microscope images were made of interlayer Josephson vortices emerging from the $a-c$ face of Tl-2201 \cite{moler1998iij} (Fig. \ref{fig:tlsci}) and Hg-1201 \cite{kirtley1998cap}. The extent of these vortices in the $a$-axis direction is set by the interlayer penetration depth $\lambda_c$. Some spreading of the vortex fields occurs as the crystal surface is approached \cite{kirtley1999mfo}, but not enough to affect the qualitative conclusions from these studies: the experimentally determined interlayer penetration depth $\lambda_c$ is an order of magnitude larger than required by the interlayer tunneling model. These conclusions are in agreement with measurements of the Josephson plasma frequency in these materials \cite{tsvetkov1998glm}.

\begin{figure}[tb]
        \begin{center}
                \includegraphics[width=10cm]{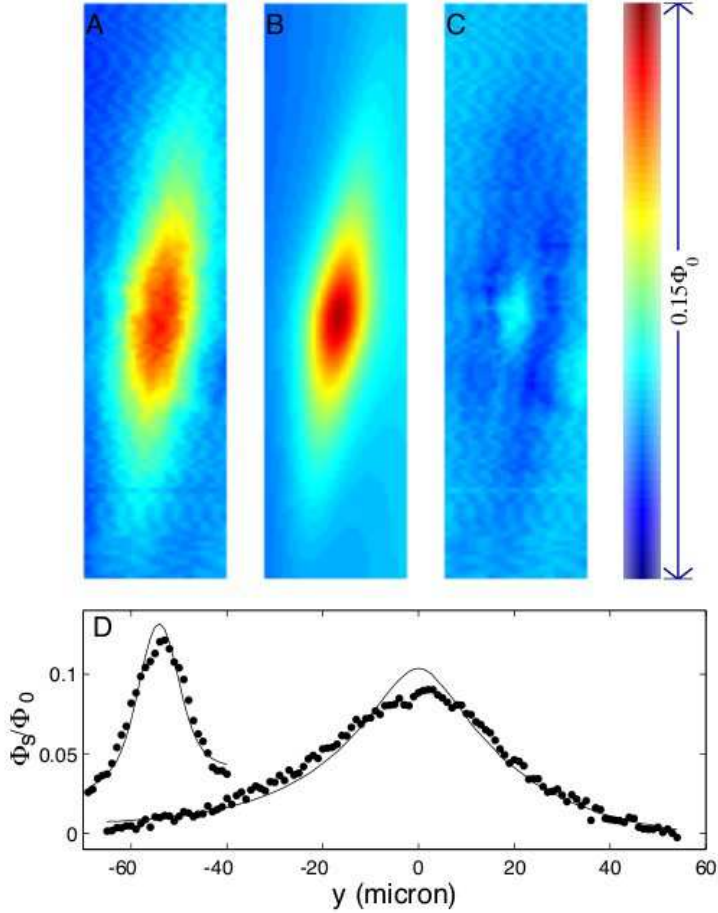}
        \end{center}
        \caption{{\bf A} Scanning SQUID microscope image of an interlayer Josephson vortex emerging from the $ac$ face of a single crystal of the single-layer cuprate superconductor Tl$_2$Ba$_2$CuO$_{6+\delta}$, imaged at 4.2K with a square pickup loop 8.2 $\mu$m on a side. {\bf B} A 2-dimensional fit to the vortex in {\bf A} with $\lambda_c$, the penetration depth perpendicular to the CuO$_2$ planes, equal to 18$\mu$m and $z_0$, the spacing between the SQUID pickup loop and the sample surface, equal to 3.6 $ \mu m$.  {\bf C} The difference between the data and the fit. {\bf D}  Cross sections perpendicular and parallel to the long axis of the vortex, offset for clarity. From K.A. Moler, J.R. Kirtley, D.G. Hinks, T.W. Li, and Ming Xu,  \href{http://www.sciencemag.org/cgi/content/abstract/279/5354/1193}{Science  {\bf 279}, 1193 (1998)}. Reprinted with permission from AAAS.}

        \label{fig:tlsci}
\end{figure}

\subsubsection{Pearl vortices}

\begin{figure}[tb]
        \begin{center}
                \includegraphics[width=10cm]{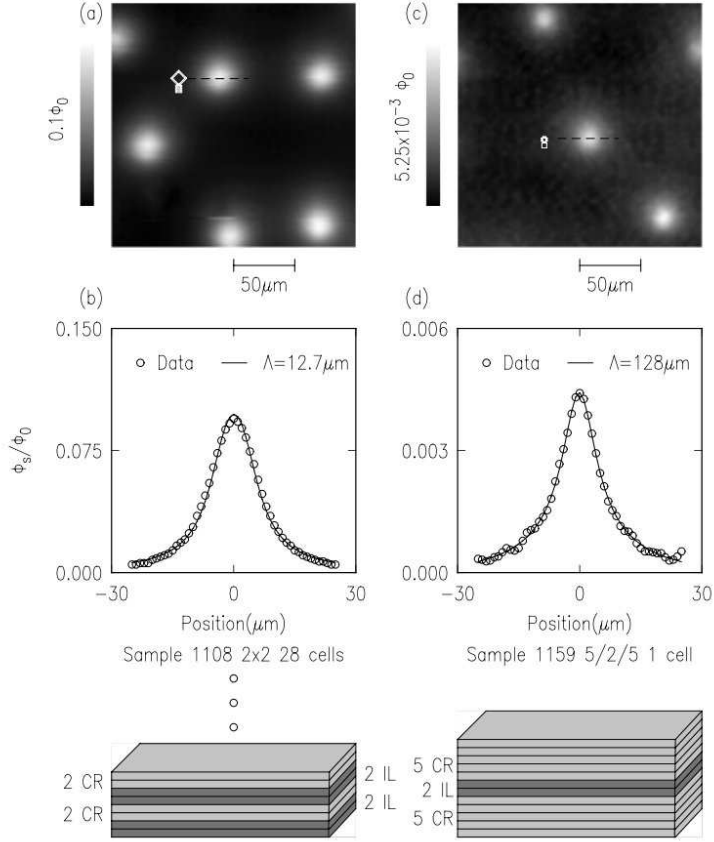}
        \end{center}
        \caption{SQUID microscope image and cross-sectional data along the positions indicated by the dashed line in (a), (c) of Pearl vortices trapped in two artificially layered (Ba$_{0.9}$Nd$_{0.1}$CuO$_{2+x}$)$_m$/(CaCuO$_2$)$_n$ superconducting films. The SQUID pickup loops were a square 7.5 $\mu$m on a side (a),(b), and an octagon 4$\mu$m on a side (c),(d) (schematics superimposed on image). The open symbols in (b),(d) are the cross-sectional data; the solid lines in (b), (d) are fits. Reprinted figure with permission from F. Tafuri, J.R. Kirtley, P.G. Medaglia, P. Orgiani, and G. Balestrino,  \href{http://prl.aps.org/abstract/PRL/v92/i15/e157006}{Phys. Rev. Lett. {\bf 92}, 157006 (2004)}. Copyright 2004 by the American Physical Society.}
        \label{fig:cbcofig1}
\end{figure}

Vortices in thin superconductors satisfying $d << \lambda_L$, where $d$ is the thickness and $\lambda_L$ is the London penetration depth, first described by Pearl \cite{pearl1964cds}, have several interesting attributes. The field strengths $H_z$ perpendicular to the films diverge as $1/r$ at distances $r <<\Lambda$, where $\Lambda=2\lambda_L^2/d$ is the Pearl length, in Pearl vortices, whereas in Abrikosov (bulk) vortices the fields diverge as $\ln(r/\lambda_L)$ \cite{kogan2001jjt}. Since in the Pearl vortex much of the vortex energy is associated with the fields outside of the superconductor, the interaction potential $V_{int}(r)$ between Pearl vortices has a long range component $V_{int} \sim \Lambda/r$ for $r >> \Lambda$ \cite{pearl1964cds}, unlike Abrikosov vortices, which have only short range interactions. Although it was originally believed that the interaction between Pearl vortices $V_{int} \sim \ln(\Lambda/r)$ for $r << \Lambda$ leads to a Berezinskii-Kosterlitz-Thouless (BKT) transition which is cut off due to screening on a scale $\Lambda$ \cite{blatter1994vht}, Kogan \cite{kogan2007ivt} reports that  the BKT transition could not happen in thin superconducting films of any size on insulating substrates because of boundary conditions at the film edges that turn the interaction into a near exponential decay. Grigorenko {\it et al.}, using SHM, observed domains of commensurate vortex patterns near rational fractional matching fields of a periodic pinning array in a thin film superconductor, which could only form when the vortex-vortex interactions are long range \cite{grigorenko2003slc}.

Figure \ref{fig:cbcofig1} shows scanning SQUID microscope images of Pearl vortices in ultra-thin [Ba$_{0.9}$Nd$_{0.1}$CuO$_{2+x}$]$_m$/[CaCuO$_2$]$_n$ (CBCO) high-temperature superconductor films. Pearl vortices can be magnetically imaged in such thin films, although the Pearl length can be as large as 1 mm, because of the strong $1/r$ divergence of the Pearl vortex magnetic fields at the vortex core: The thin-film limit for the two-dimensional Fourier transform of the $z$-component of the field from an isolated vortex trapped in a thin film is given by \cite{pearl1964cds,kogan1994mfv}:
\begin{equation}
B_z(k,z)=\frac{\Phi_0 e^{-k z}}{1+k \Lambda},
\label{eq:thinvort}
\end{equation}
where $z$ is the height above the film, $k=\sqrt{k_x^2+k_y^2}$, $\Lambda=2\lambda_{ab}^2/d$, $\lambda_{ab}$ is the in-plane penetration depth, $d$ is the film thickness and $\Phi_0 = h/2e$. 

The open circles in Fig. \ref{fig:cbcofig1}(b,d) are cross-sections through the data of Fig. \ref{fig:cbcofig1}(a,c). The solid lines are fits to the data using Eq. \ref{eq:thinvort}, using the Pearl length $\Lambda$ as a fitting parameter. The in-plane London penetration depth $\lambda_{ab}$ of these thin films can be inferred from $\Lambda$ if $d$ is known.

\subsubsection{Mesoscopic superconductors}

\begin{figure}[tb]
        \begin{center}
                \includegraphics[width=10cm]{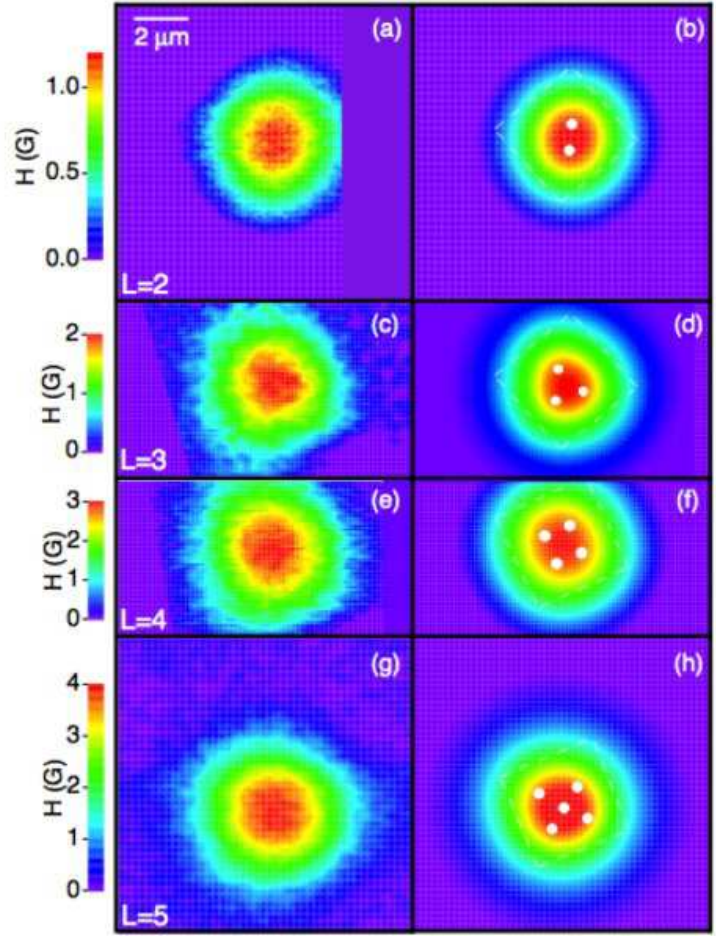}
        \end{center}
        \caption{ Scanning Hall probe microscope images of a 4 $\mu$m $\times$ 4$\mu$m square Pb film at applied fields of (a) 4.6 G, (c) 6.6 G, (e) 8.6 G,  and (g) 10.6 G. [(b), (d), (f), and (h)] The results of 2D monopole fits to (a), (c), (e), and (g) respectively. The broken lines denote the shape of the square. The locations of the vortices are shown as white points. Reprinted figure with permission from T. Nishio, Q. Chen, W. Gillijns, K. De Keyser, K. Vervaeke, and V.V. Moshchalkov,  \href{http://prb.aps.org/abstract/PRB/v77/i1/e012502}{Phys. Rev. B {\bf 77}, 012502 (2008)}. Copyright 2008 by the American Physical Society.} 
         \label{fig:mesoscopic_square}
\end{figure}

Vortices in superconductors with dimensions comparable to the superconducting coherence length $\xi$ have been predicted to nucleate spontaneously in a number of interesting configurations, such as shells \cite{lozovik1998ebs}, giant vortices \cite{fink1966mis}, and anti-vortices, so as to preserve the symmetry of the sample \cite{chibotaru2000sif,chibotaru2001ven}. Vortex trapping patterns in mesoscopic superconducting samples have been studied using Bitter decoration \cite{grigorieva2006dov}, scanning SQUID microscopy \cite{nishio2004pvi}, ballistic Hall magnetometry \cite{geim1997bhm}, transport \cite{moshchalkov1995est}, and Hall bar microscopy. An example of Hall bar microscopy results is shown in Figure \ref{fig:mesoscopic_square}. Nishio {\it et. al.} \cite{nishio2008shp} report results in agreement with Ginzburg-Landau calculations for temperatures considerably lower than the superconducting transition temperature.

\subsubsection{Sr$_2$RuO$_4$}

\begin{figure}[tb]
        \begin{center}
                \includegraphics[width=14cm]{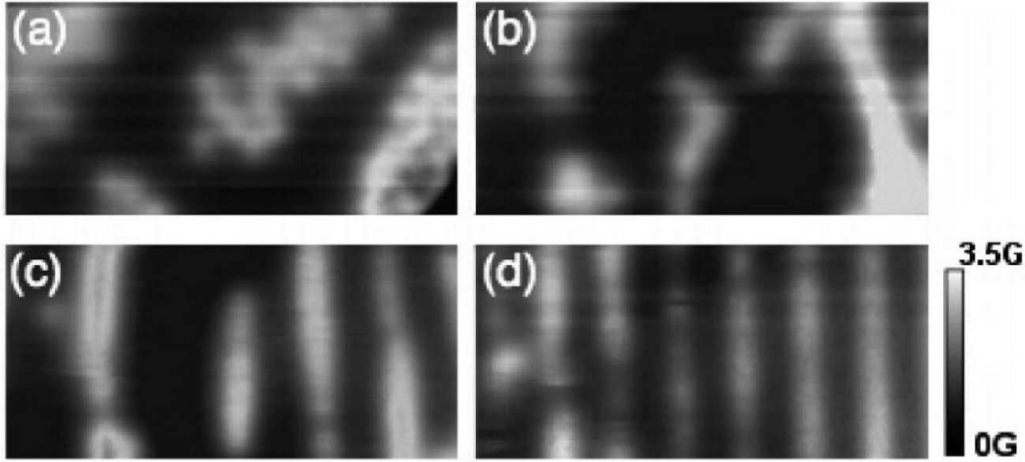}
        \end{center}
        \caption{Scanning SQUID microscope images of flux domains in Sr$_2$RuO$_4$ at T=0.36 K after field cooling at various fields. In all cases, the magnetic field amplitude applied along the $c$ axis (H$_\perp$) was kept constant at 2 G while the in-plane field (H$_{ab}$) was (a) 0 G, (b) 5 G, (c) 10 G, (d) 50 G. The imaging area is 31 $\mu$m$\times$ 17 $\mu$m. The field scale in G is shown on the right; dark regions are superconducting vortex-free regions. Reprinted figure with permission from V.O. Dolocan, C. Veauvy, F. Servant, P. Lejay, K. Hasselbach. Y. Liu, and D. Mailly,  \href{http://prl.aps.org/abstract/PRL/v95/i9/e097004}{Phys. Rev. Lett.  {\bf 95}, 097004 (2005)}. Copyright 2005 by the American Physical Society.}
        \label{fig:sr2ruo4}
\end{figure}

Sr$_2$RuO$_4$, with the same crystal structure as La$_2$CuO$_4$, the parent compound for the first class of cuprate perovskite compounds shown to exhibit high-temperature superconductivity \cite{bednorz1986pht}, was discovered in 1994 to be superconducting at about 1K \cite{maeno1994slp}. It is thought to be a $p_x\pm ip_y$-wave, triplet pairing superconductor with broken time-reversal symmetry. One of the consequences of this pairing symmetry state is spontaneously generated supercurrents, and consequent large magnetic fields at surfaces and boundaries between $p_x+ ip_y$ and $p_x - i p_y$ domains \cite{matsumoto1999qsn}. Although there is evidence for characteristic magnetic fields of 0.5G in Sr$_2$RuO$_4$ from muon spin resonance experiments \cite{luke1998trs}, and evidence for broken time-reversal symmetry at the surface of Sr$_2$RuO$_4$ has been obtained using Sagnac interferometry \cite{xia2008pke},
no evidence for spontaneously generated magnetic fields at the surfaces of Sr$_2$RuO$_4$ samples has been found in scanning Hall bar and scanning SQUID experiments, with much higher sensitivity than the magnetic fields originally predicted \cite{bjornsson2005brs,kirtley2007uls,hicks2010lsr}. A possible explanation for this failure to observe spontaneous magnetization directly is that the $p_x\pm i p_y$ domains are small, so that the fields from closely spaced domain boundaries nearly cancel each other. The experimental information on $p_x\pm i p_y$ domain sizes is not consistent, with the first phase sensitive pairing symmetry experiments on Sr$_2$RuO$_4$ being consistent with domain sizes of order 1 mm \cite{nelson2004ops}, while later phase-sensitive results are consistent with domain sizes 1$\mu m$ or less \cite{kidwingira2006dso}, and the Sagnac interferometry being consistent with domain sizes intermediate between these two sizes \cite{xia2008pke}. Recently Raghu {\it et al.} have suggested that the superconductivity in Sr$_2$RuO$_4$ arises primarily in quasi-one-dimensional bands, and is therefore not expected to generate experimentally detectable edge currents \cite{raghu2010hqo}.

Dolocan {\it et al.} \cite{dolocan2005ovc,dolocan2006ots} have magnetically imaged single crystals of Sr$_2$RuO$_4$ using a scanning $\mu$SQUID mounted on a tuning fork in a dilution refrigerator. This allowed combined magnetic and topographic information on a length scale of 1 micron, at temperatures below 400 mK. They found coalescence of vortices and the formation of flux domains (see Fig. \ref{fig:sr2ruo4}). The formation of lines of vortices in a field tilted relative to the crystalline $c$-axis \cite{grigorenko2001odc} can be attributed in this case to the large anisotropy in Sr$_2$RuO$_4$, resulting in an attractive interaction between vortices in a plane defined by the $c$-axis and the magnetic field direction \cite{dolocan2006ots}. The vortex coalescence observed when the applied magnetic field is parallel to the $c$ axis may be due to weak intrinsic pinning at domain walls, and the existence of a mechanism for bringing vortices together that overcomes the conventional repulsive vortex-vortex interaction.

\subsubsection{Magnetic superconductors}

\begin{figure}[tb]
        \begin{center}
                \includegraphics[width=10cm]{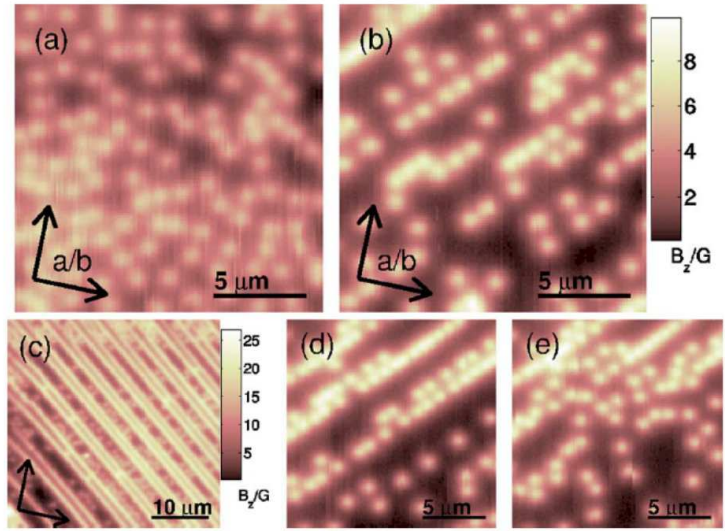}
        \end{center}
        \caption{Scanning Hall bar image of a single crystal sample of ErNi$_2$B$_2$C. After cooling below T$_c$ in a weak field [2.4 Oe in (a), (b), (d), (e), 18 Oe in (c)], a vortex distribution consistent with random pinning is observed on the $ab$ face [(a) 7.3 K]. Upon reducing the temperature below T$_N$, the vortices spontaneously organize along twin domain walls along the [110] [(b) 5.3 K] or [1${\bar 1}$0] [(c) 4.2 K] direction. The pattern gradually disappears as the temperature is raised again [(d) 5.7 K, (e) 6.0 K, different cycle than (a) and (b)]. Reprinted figure with permission from H. Bluhm, S.E. Sebastian, J.W. Guikema, I.R. Fisher, and K.A. Moler,  \href{http://prb.aps.org/abstract/PRB/v73/i1/e014514}{Phys. Rev. B {\bf 73}, 014514 (2006)}. Copyright 2006 by the American Physical Society.} 
         \label{fig:bluhm_fig_2}
\end{figure}

Magnetic order and superconductivity are competing orders since the Meissner state excludes a magnetic field from the bulk and superconductivity is destroyed at sufficiently high fields. However, superconductivity and magnetism can coexist \cite{saxena2000sbi,aoki2001csf,huy2007sbw,slooten2009esn} if the orientation of the local magnetic moments varies on a length scale shorter than the superconducting penetration depth $\lambda$, or if the field generated by the magnetization is carried by a so-called spontaneous vortex lattice \cite{tachiki1980siv}. The former order has been observed in ErRh$_4$B$_4$ and HoMo$_6$S$_8$ \cite{moncton1980omf,lynn1981cfs}, and indirect evidence for the latter has been reported in UCoGe \cite{ohta2010mcf}. Several Hall bar studies have been performed on artificial ferromagnetic-superconducting hybrid structures \cite{fritzsche2009vvm,vanbael2001lof,menghini2009dvm,silevitch2001iam}. Fig. \ref{fig:bluhm_fig_2} \cite{bluhm2006shp} shows scanning Hall bar microscopy measurements of ErNi$_2$B$_2$C, which has a superconducting T$_c$ of $\approx$ 11 K, becomes antiferromagnetic below T$_N \approx$ 6 K, and exhibits weak ferromagnetism below T$_{WFM} \approx$ 2.3 K \cite{canfield1996pce}. Bluhm {\it et al.} \cite{bluhm2006shp} found that in this material the superconducting vortices spontaneously rearranged to pin on twin boundaries upon cooling through the antiferromagnetic transition temperature, and that a weak random magnetic signal appears in the ferromagnetic phase.

\subsection{Fluxoid dynamics}

\subsubsection{Limits on spin-charge separation}

\label{sec:spin_charge}

\begin{figure}[tb]
        \begin{center}
                \includegraphics[width=10cm]{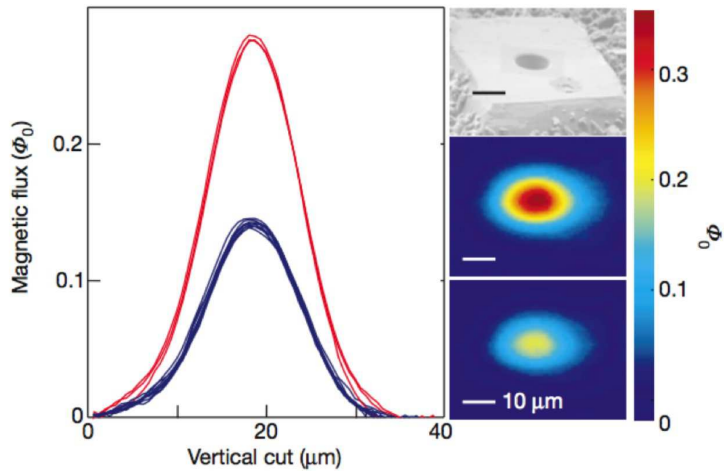}
        \end{center}
        \caption{Cross-sections and images of single (bottom right image, blue cross-section) and double (middle right image, red cross section) flux quanta trapped in a 50$\mu$m $\times$ 50$\mu$m square of single-crystal YBa$_2$Cu$_3$O$_{6.35}$ (T$_c$=6.0K) with a 10 $\mu$m hole drilled using a focussed ion beam (top right image). When the sample was heated to 5.6K for 1 sec and re-cooled to 2K, these flux quanta escaped with no sign of the vortex memory associated with visons. Reprinted figure with permission from Macmillan Publishing Ltd: D.A. Bonn, J.C. Wynn, B.W. Gardner, Y.-J. Lin, R. Liang, W.N. Hardy, J.R. Kirtley and K.A. Moler,  \href{http://www.nature.com/nature/journal/v414/n6866/abs/414887a.html}{Nature  {\bf 414}, 887 (2001)}. Copyright 2001.}

        \label{fig:bonnnat}
\end{figure}

One explanation for the peculiar normal state properties and high superconducting transition temperatures of the high-T$_c$ cuprate perovskites is that there exists a new state of matter in which the elementary excitations are not electron-like, as in conventional metals, but rather the carriers ``fractionalize" into ``spinons", chargeless spin 1/2 fermions, and ``chargons", bosons with charge $+e$ \cite{anderson1987rvb}. Senthil and Fisher proposed a model for such a separation in two dimensions with sharp experimental tests \cite{senthil2001fic}: in conventional superconductors a Cooper pair with charge $2e$ circling around a vortex with $h/2e$ of total magnetic flux experiences a phase shift of 2$\pi$. However, a chargon would pick up only a phase shift of $\pi$. In order for the chargon wave function to remain single-valued, it must be accompanied by a ``vison" that provides an additional $\pi$ phase shift. In the Senthil-Fisher test, a superconducting ring with flux trapped in it is warmed through the superconducting transition. If the number of vortices in the ring is odd, then it will also contain a vison that could persist above T$_c$. If the ring is then recooled in zero field before the vison escapes, it will spontaneously generate a single flux quantum with random sign. On the other hand, if the number of vortices in the ring is even, no such ``vortex memory" effect would exist. Bonn {\it et al.} \cite{bonn2001lsc} (Fig. \ref{fig:bonnnat}) tested for this effect in highly underdoped single crystals of YBa$_2$Cu$_3$O$_{6+x}$. These crystals have a very sharp but low transition temperature favorable for such a test. They found no evidence for a vortex memory effect, and were able to put stringent upper limits on the energy associated with a vison. Some models in which superconductivity results from spin-charge separation predicts $h/e$ fluxoids in materials with low superfluid density \cite{sachdev1992shv,nagaosa1992glt,nagaosa1994e2q}. Wynn {\it et al.} \cite{wynn2001lsc}, using scanning SQUID and Hall bar microscopies, found no evidence for $h/e$ vortices in strongly underdoped YBa$_2$Cu$_3$O$_{6+x}$ crystals, also placing limits on spin-charge separation in these materials.

\subsubsection{Superconducting rings}

\begin{figure}[tb]
        \begin{center}
                \includegraphics[width=10cm]{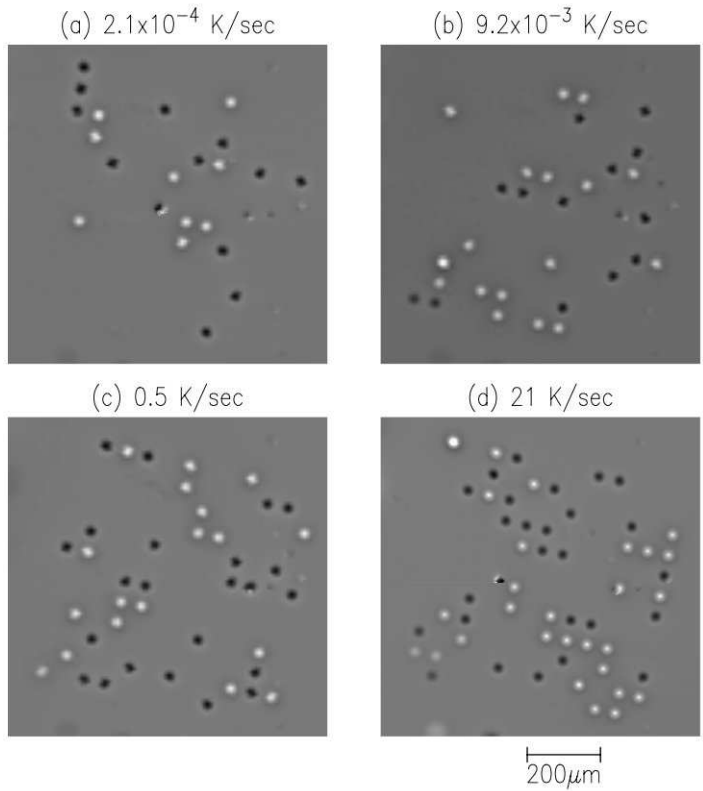}
        \end{center}
        \caption{SQUID microscope images of a 12 $\times$ 12 array of 20 $\mu$m inside diameter, 30 $\mu$m outside diameter thin film rings of Mo$_3$Si, cooled in zero field through the superconducting T$_c$ at various rates. The resulting ring vortex numbers were consistent with thermally activated spontaneous fluxoid formation. Reprinted figure with permission from J.R. Kirtley, C.C. Tsuei, and F. Tafuri,  \href{http://prl.aps.org/abstract/PRL/v90/i25/e257001}{Phys. Rev. Lett. {\bf 90}, 257001 (2003)}. Copyright 2003 by the American Physical Society.}
        \label{fig:cosmprl1}
\end{figure}

As we have seen in previous sections, the dynamics of fluxoids in a ring configuration are of interest as a model for the Ising spin system \cite{davidovic1996cda,davidovic1997mcg,kirtley2005aoa} \ref{sec:ring_arrays}, and for placing limits on spin-charge separation in underdoped cuprate superconductors \cite{bonn2001lsc} \ref{sec:spin_charge}. Kirtley {\it et al.} studied fluxoid dynamics in photolithographically patterned thin film rings of the underdoped high-temperature superconductor Bi$_2$Sr$_2$CaCu$_2$O$_{8+\delta}$ using a scanning SQUID microscope, and concluded that their results could be understood in terms of thermally activated nucleation of a Pearl vortex in, and transport of the Pearl vortex across, the ring wall \cite{kirtley2003fds}. 

Fluxoid dynamics in superconducting rings may also provide clues to physics on an entirely different scale: the early development of the universe. Immediately after the Big Bang a fundamental symmetry related the particles that mediated the electromagnetic, strong, and weak forces. However, as the universe cooled down this symmetry was broken in such a way that, for example, the $W$ and $Z$ bosons, which mediate the weak interaction, have mass, while photons, which mediate the electromagnetic force, do not. Kibble proposed an intuitive picture of this symmetry breaking of the early universe \cite{kibble1976tcd,kibble2007ptd}, with the underlying idea that causality governed the number of defects because a finite time delay is needed for information to be transferred between different regions of a system. Zurek pointed out that Kibble's ideas could be tested by studying vortices in superfluids and superconductors \cite{zurek1985ces}. As a superconductor or superfluid cools through its transition temperature different regions can nucleate into a state with the same order parameter amplitude $\psi$ but different phases $\phi$. This will produce ``kinks" or ``defects" where $\phi$ would eventually (at low temperature) change by 2$\pi$. These kinks can be removed by annihilation with anti-kinks if the coherence length $\xi$ is long enough and the cooling rate is sufficiently slow. However, as the system cools this annihilation process proceeds less rapidly, until at some point the kinks ``freeze-in" because part of the system cannot communicate with other parts quickly enough. In a superconducting ring a phase shift of 2$\pi$ around the ring corresponds to a fluxoid trapped in the ring, so that the process of forming and freezing in kinks is also called spontaneous fluxoid generation. Kibble and Zurek showed that causality implies that the probability $P$ of finding a spontaneous fluxoid in the ring (when cooled in zero external field) depends on the cooling rate $\tau_Q^{-1}$ as $P \propto (\tau_Q/\tau_0)^{-\sigma}$, where $\tau_0$ is a characteristic time and $\sigma$ depends on how the correlation length $\xi$ and the relaxation time $\tau$ vary with temperature close to the transition. For a superconducting ring and assuming a mean-field approximation $\sigma=1/4$.

Tests of the Kibble-Zurek prediction have been made by looking for vortices in superfluid $^4He$ \cite{hendry1994gds,dodd1998nvf} and $^3He$ \cite{bauerle1996lsc,ruutu1996vfn}, in superconducting films \cite{carmi1999spn,golubchik2009cbh}, and superconducting rings interrupted by Josephson junctions \cite{carmi2000osf,kavoussanaki2000tkz,monaco2002zkd}. However, verification of the ideas of Kibble and Zurek using arguably the simplest system, superconducting rings, were not made until recently. Kirtley, Tsuei and Tafuri \cite{kirtley2003tas} (Fig. \ref{fig:cosmprl1}) used a SQUID microscope to image an array of rings after repeated cooling in various magnetic fields and cooling rates to determine the probability of spontaneous fluxoid generation. They showed that a second mechanism prevailed, in which the final density of fluxoids depended on a balance between thermal generation and the relaxation rates of fluxoids \cite{hindmarsh2000dfl,ghinovker2001smf,donaire2007svf}. They argued that the thermal fluctuation mechanism is complementary to the ``causal" mechanism and should be considered in attempts to understand phase transitions, both in the laboratory and in the early universe. The Kirtley {\it et al.} experiments \cite{kirtley2003tas} were done on rings with parameters which favored the thermal activation mechanism over the Kibble-Zurek (causal) mechanism. Recently Monaco {\it et al.} \cite{monaco2009sff}  have studied superconducting rings with parameters more favorable to the causal mechanism, using a geometry in which the flux state of a single ring can be manipulated and sensed rapidly, and find results consistent with the Kibble-Zurek prediction, taking into account the fact that the ring's circumference was much smaller than the coherence length at the temperature at which the fluctuations were frozen in.

\subsubsection{High-T$_c$ grain boundaries}

\begin{figure}[tb]
        \begin{center}
                \includegraphics[width=10cm]{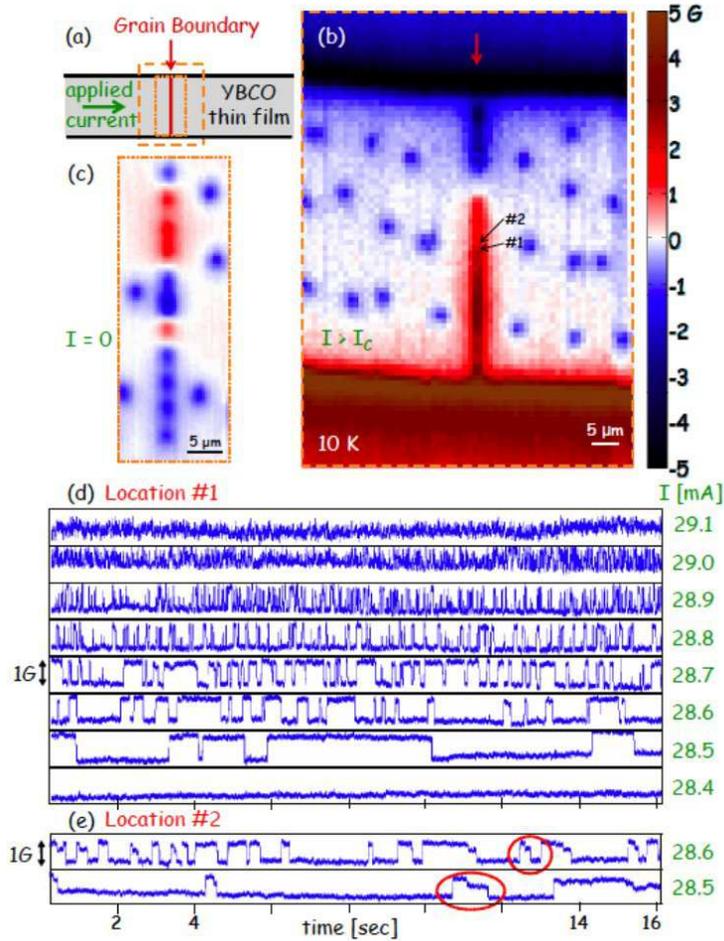}
        \end{center}
        \caption{Scanning Hall bar measurements of vortex noise in a 24$^o$ symmetric YBCO  grain boundary (GB). (a) Sketch of the sample. The bridge width is 50 $\mu$m. (b,c) Magnetic image of the GB area (b) at 10 K and applied current I = 29 mA and (c) at 10 K after thermal cycling to 70 K with I = 0, showing individual vortices (red) and antivortices (blue). The regions imaged are outlined in orange in (a). (d,e) The Hall probe signals {\it vs.} time at 10 K for different currents applied to the GB measured at two positions marked in Fig. 1b. The switching events are vortices passing under the probe. Red circles indicate three-level switching events. Reprinted figure with permission from B. Kalisky, J.R. Kirtley, E.A. Nowadnick, R.B. Dinner, E. Zeldov, Ariando, S. Wenderich, H. Hilgenkamp, D.M. Feldmann, and K.A. Moler,  \href{http://apl.aip.org/applab/v94/i20/p202504_s1}{Appl. Phys. Lett. {\bf 94}, 202504 (2009)}. Copyright 2009 by the American Institute of Physics.} 
         \label{fig:gbjnoise}
\end{figure}

Grain boundaries in the cuprate high-T$_c$ superconductors are widely used for devices and fundamental studies, and govern the transport properties of superconductors with technological applications \cite{hilgenkamp1992gbh}. It is therefore important to understand the mechanism of dissipation in transport across grain boundaries. Transport across grain boundaries is usually studied with a voltage threshold on the order of nanovolts. At these voltages millions of vortices traverse the boundary per second. Much smaller voltages can be measured by magnetically imaging the vortices in the grain boundaries. Kalisky {\it et al.} \cite{kalisky2009dsv} performed such measurements on grain boundaries produced by epitaxial growth of the cuprate high-T$_c$ superconductor YBa$_2$Cu$_3$O$_{7-\delta}$ on bicrystals of SrTiO$_3$ using a large scanning area Hall bar microscope \cite{dinner2005csh}. They observed (Figure \ref{fig:gbjnoise}) telegraph noise in the Hall bar signal when the sensor was directly above the grain boundary, which they attributed to the motion of vortices. They inferred the voltage across the grain boundary from the frequency of the telegraph noise, with voltage sensitivity as small as 2$\times$10$^{-16}$V. The dependence of the inferred grain boundary voltage on current followed either an exponential $V \propto e^{mI/I_c}$ or a power law $V \propto I^n$ dependence with high values of $m$ or $n$. These results were qualitatively different from grain boundary transport measurements at higher voltages \cite{hilgenkamp1992gbh,dimos1988odg,heinig1999swc}, and could not be fully explained using existing models.

\subsection{Local susceptibility measurements}

\subsubsection{Superconducting fluctuations}

\begin{figure}[tb]
        \begin{center}
                \includegraphics[width=14cm]{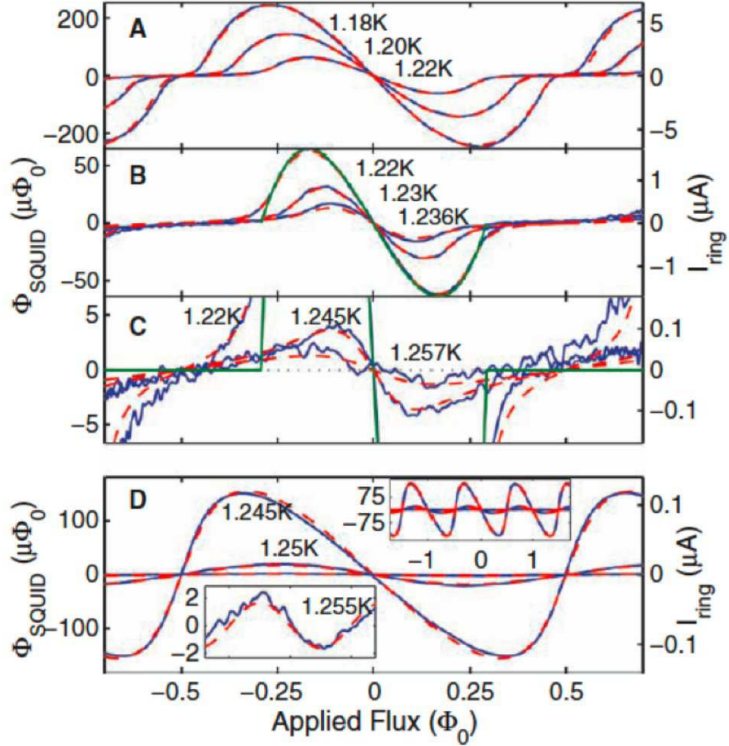}
        \end{center}
        \caption{SQUID signal (left axis) and ring current (right axis) as a function of applied flux $\Phi_a$ for two aluminum rings, both with thickness d = 60 nm and annulus width w = 110 nm. The fluctuation theory (dashed red) was fit to the data (blue) through the dependence of $I_{\rm ring}$ on $\Phi_a$ and temperature. (A to C) ring radius R = 0.35 $\mu$m, fitted T$_c$($\Phi_a$=0) = 1.247 K, and $\gamma$ = 0.075. The parameter $\gamma$ characterizes the size of the ring. The green line is the theoretical mean field response for T = 1.22 K and shows the characteristic Little-Parks line shape, in which the ring is not superconducting near $\Phi_a$ = $\Phi_0/2$. The excess persistent current in this region indicates the large fluctuations in the Little-Parks regime. (D) R = 2 $\mu$m, fitted T$_c$($\Phi_a$=0) = 1.252 K, and $\gamma$ = 13. From N.C. Koshnick, H. Bluhm, M.E. Huber, and K.A. Moler,  \href{http://www.sciencemag.org/cgi/content/abstract/318/5855/1440}{Science {\bf 318}, 1440 (2007)}. Reprinted with permission from AAAS.}
        \label{fig:fluctuations}
\end{figure}

Experimental knowledge of superconducting fluctuations in one dimension has largely been derived from transport measurements \cite{lau2001qps}, which require electrical contacts and an externally applied current. The SQUID susceptometer, along with a ring geometry for the superconductor, allows the study of fluctuation effects, without contacts, in isolated, quasi one-dimensional rings in the temperature range where the circumference is comparable to the temperature-dependent Ginzburg-Landau coherence length $\xi(T)$.  In such measurements, an example of which is shown in Fig. \ref{fig:fluctuations} \cite{koshnick2007fsm}, a magnetic field is applied to the ring using a single-turn field coil, co-planar and concentric with the pickup loop, both of which are integrated into the SQUID sensor \cite{gardner2001ssq}. The response of the ring to this applied flux is measured by the pickup loop. The SQUID sensor is in a carefully balanced gradiometer configuration, such that it is insensitive to the fields applied by the field coil \cite{huber2008gms}. Nevertheless, further background subtraction, by comparing the mutual inductance between the field coil and SQUID with the pickup loop close to the ring with that with the pickup loop at a distance, is required to measure the response field due to the ring, which can be 7 orders of magnitude smaller than the applied field. A scanned sensor allows multiple rings to be measured in a single cooldown. The experimental results agree with a full numeric solution of thermal fluctuations in a Ginzburg-Landau framework that includes non-Gaussian effects \cite{von1992fpp,scalapino1972smo} for all of the rings for which calculations were numerically tractable. This is in contrast to previous work on a single Al ring \cite{zhang1997sms}, which disagreed strongly with theory.

Similar techniques have allowed the measurement of spontaneous persistent currents in normal metal rings \cite{bluhm2009pcn}.

\subsubsection{Stripes}

\begin{figure}[tb]
        \begin{center}
                \includegraphics[width=14cm]{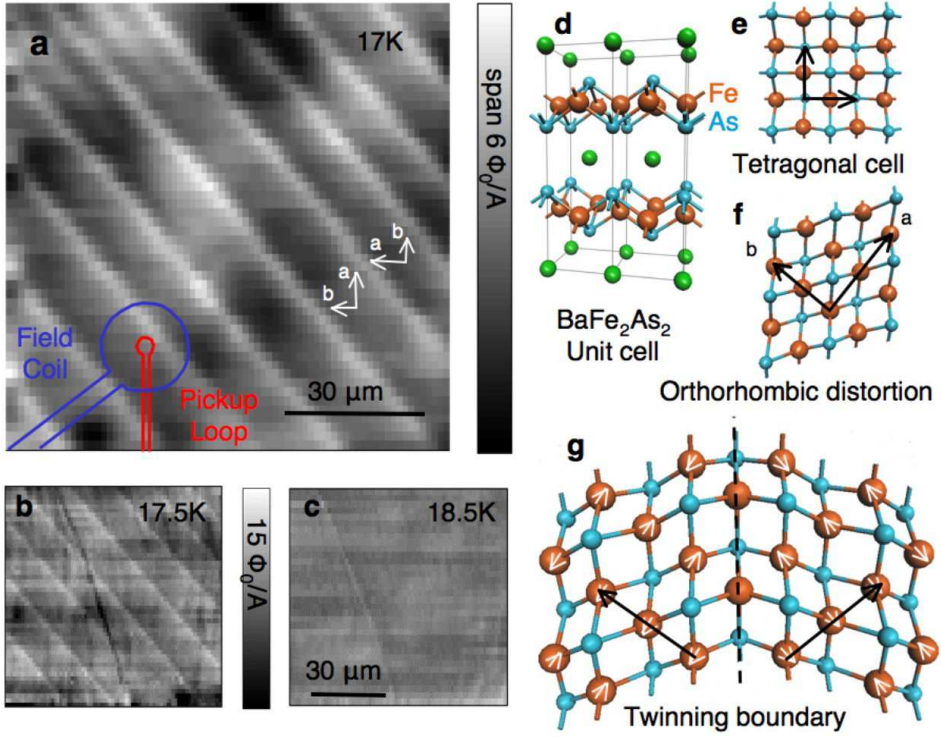}
        \end{center}
        \caption{Local susceptibility image in underdoped Ba(Fe$_{1-x}$Co$_x$)$_2$As$_2$, indicating increased diamagnetic shielding on twin boundaries. (a) Local diamagnetic susceptibility, at T=17 K, of the $ab$ face of a sample with $x$=0.051 and T$_c$=18.25 K, showing stripes of enhanced diamagnetic response (white). In addition there is a mottled background associated with local T$_c$ variations that becomes more pronounced as T $\rightarrow$ T$_c$. Overlay: sketch of the scanning SQUID sensor. The size of the pickup loop sets the spatial resolution of the susceptibility images. (b) and (c) Images of the same region at (b) T=17.5 K and (c) at T=18.5 K show that the stripes disappear above T$_c$. A topographic feature (scratch) appears in (b) and (c). (d) Crystal structure of a unit cell of the parent compound BaFe$_2$As$_2$. (e) Top view of the FeAs layer with tetragonal symmetry, and (f) an exaggerated view of the orthorhombic distortion that occurs at low temperatures.  (g) Schematic of a possible arrangement of spins across a twin boundary in the anti-ferromagnetic state. Reprinted figure with permission from B. Kalisky, J.R. Kirtley, J.G. Analytis, J.-H. Chu, A. Vailionis, I.R. Fisher, and K.A. Moler,  \href{http://prb.aps.org/abstract/PRB/v81/i18/e184513}{Phys. Rev. B  {\bf 81}, 184513 (2010)}. Copyright 2010 by the American Physical Society.}
        \label{fig:stripes_Figure_1}
\end{figure}

The pnictide class of superconductors \cite{takahashi2008saf} have the highest critical temperatures (57K) observed for a non-cuprate superconductor, multiband Fermi surfaces, and at certain doping levels a paramagnetic to antiferromagnetic as well as a tetragonal to orthorhombic transition above the superconducting transition temperature. In Ba(Fe$_{1-x}$Co$_x$)$_2$As$_2$ (Co doped Ba-122) and other members of the 122 family (AFe$_2$As$_2$ with A=Ca, Sr, and Ba), doping causes the spin-density-wave transition temperature and the structural transition temperature to decrease \cite{chu2009dpd,ni2008ecs} falling to zero at or near the doping where the highest T$_c$ occurs, suggesting the importance of lattice changes in determining transport properties. Both experiment \cite{kimber2009sbs,liu2009lii,pickett2009ibs} and theory \cite{yildirim2009scf} have suggested a close relationship between structural and magnetic properties, leading some authors to describe the lattice and spin density-wave transition by a single order parameter \cite{fang2008ten,xu2008iso}. In addition, structural strain appears to play a significant role in the superconductivity. For example, in the 122 compounds, small amounts of non-hydrostatic pressure can induce superconductivity \cite{colombier2009cpd,goldman2009lcq,kreyssig2008piv,yu2009ass,alireza2009sut}. In addition, there is evidence that the structural perfection of the Fe-As tetrahedron is important for the high critical temperatures observed in the Fe pnictides \cite{zhao2008smp,lee2008esp}.

Kalisky {\it et al.} \cite{kalisky2010sid}  have shown that Co doped Ba-122 has stripes of enhanced susceptibility below the superconducting transition temperature (see Fig. \ref{fig:stripes_Figure_1}). These stripes are resolution limited using a SQUID pickup loop with an effective diameter of 4$\mu$m, and are believed to be associated with boundaries between twins- crystallites with their in-plane $a,b$ axes rotated by 90$^o$. The amplitude of the susceptibility stripes becomes larger as the superconducting transition temperature is approached from below. Since the susceptibility signal for a homogeneous superconductor becomes larger as the penetration depth becomes shorter and the superfluid density becomes larger, it seems reasonable to associate the susceptibility stripes with shorter penetration depths and enhanced superfluid densities on the twin boundaries. Twin boundaries have been associated with enhanced superfluid density in conventional superconductors \cite{khlyustikov1987tps}, but the influence of twin planes on superconductivity in, for example, the cuprates is less clear \cite{fang1988ptb,dalidovich2008sdn,sonier1997mfl,abrikosov1989ptt,deutscher1987osg}.

It is difficult to calculate analytically the effect of a plane of reduced penetration depth (enhanced superfluid density) on the susceptibility observed in scanning SQUID measurements. Kirtley {\it et al.} \cite{kirtley2010mrb} have done a finite-element calculation in the appropriate geometry. Because of the uncertainty in the width of the region with enhanced superfluid density, it is also difficult to estimate the size of the enhancement in the Cooper pair density from the scanning susceptibility measurements. Nevertheless, Kirtley {\it et al.} estimate an enhancement in the two-dimensional Cooper pair density in the range between 10$^{19}$ and 10$^{20}$ m$^{-2}$. For comparison, the two-dimensional electron liquid \cite{breitschaft2010tde} at the LaAlO$_3$-SiTiO$_3$ interface, which exhibits superconductivity at about 0.2K \cite{reyren2007sib}, has a carrier concentration of about 10$^{17}$ m$^{-2}$. The temperature dependence of the stripe amplitude can be best fit by assuming that the twin boundary has a different critical temperature than the bulk, although no stripes were observed above the bulk T$_c$.

Kalisky {\it et al.} also observed that vortices tended to avoid pinning on the twin boundaries, and that they were difficult to drag over the twin boundaries using either a SQUID susceptometer sensor or a magnetic force microscope tip \cite{kirtley2010mrb}. The enhancement of supercurrent density, as well as the barrier to vortex motion presented by the twin boundaries, provide two mechanisms for enhanced critical currents in twinned samples. Prozorov {\it et al.} \cite{prozorov2009ips} have noted an enhancement of the critical current of slightly underdoped single crystals of Ba(Fe$_{1-x}$Co$_x$)$_2$As$_2$, which they associate with twinning.

\subsubsection{Spinlike susceptibility}

There appears to be a component of $1/f^\alpha$ ($\alpha \sim$ 1) noise in superconducting devices, which is apparently due to magnetic flux noise, and is remarkably universal \cite{wellstood1987lfn,koch1983fnt}. Recently Koch, DiVincenzo, and Clarke (KDC) have proposed that this noise is due to unpaired, metastable spin states \cite{koch2007mff}. The proposed spin states are bistable, with a broad distribution of activation energies for changing their spin orientation. When the spin state changes it changes the amount of magnetic flux coupling into the superconducting device. Each contributes a Lorentzian with a particular cutoff frequency to the total noise; adding up a large number of Lorentzians with a broad distribution of cutoff frequencies results in noise with $1/f$ distribution \cite{dutta1979esn}. KDC estimate that a density of unpaired spin states of about 5$\times$10$^{17}m^{-2}$ is required to explain universal $1/f$ noise. 

Bluhm {\it et al.} \cite{bluhm2009ssm} have found a signal in scanning SQUID susceptometry measurements at low temperatures in a number of materials, both metals and insulators, even including Au, that has a paramagnetic response with a temperature dependence consistent with unpaired spins. This susceptibility has a component that is out of phase with the applied field, implying that it could contribute to $1/f$-like magnetic noise. The density of these spin states is estimated to be in the range of 10$^{17}$m$^{-2}$, consistent with the KDC estimate. This implies that scanning SQUID susceptometry could be used as a diagnostic for determining which materials and processes contribute most strongly to $1/f$ noise in superconducting devices. 

\subsection{Penetration depths}

Penetration depths have been inferred from fitting the magnetic images of superconducting vortices for a number of years \cite{chang1992shp,chang1992shp2,tafuri2004mip,bending1999lmp,kirtley1999vst,nazaretski2009dmp,luan2010lmp}. An alternate means is to measure the mutual inductance between the SQUID pickup loop and a field coil integrated into the sensor in SSM \cite{gardner2001ssq} (Fig. \ref{fig:cbcofig2}). The 2-dimensional Fourier transform of the $z$-component of the response field in the pickup loop, with a current $I$ in a circular ring of radius $R$ oriented parallel to, and a height $z$ above a homogeneous bulk superconductor with penetration depth $\lambda$ is given by \cite{kogan2003mra}
\begin{equation}
B_z(k)=-\pi\mu_0 I R \left ( \frac{q-k}{q+k} \right )  J_1(k R) e^{-2 k z},
\label{eq:bzbulk}
\end{equation}
where $q=\sqrt{k^2+1/\lambda^2}$, and $J_1$ is the order 1 Bessel function of the first kind. Similarly the response field for a thin film superconducting sample with Pearl length $\Lambda$ is given by \cite{kogan2003mra}
\begin{equation}
B_z(k)=-\pi \mu_0 I R J_1(kR) \frac{e^{-2kz}}{1+k \Lambda}.
\label{eq:kogsusc}
\end{equation}
To a good approximation, if $z>>\lambda$ the magnetic fields resulting from the screening of the field coil fields by the sample act as if they are due to an image coil spaced by a distance $2h_{eff}=2(z+\lambda)$ from the real coil, where $z$ is the physical spacing between the sample surface and the field coil. A SQUID sensor with a single circular field coil of radius $R$ and a co-planar, coaxial circular pickup loop of radius $a$ oriented parallel to a homogeneous bulk sample surface will have a mutual inductance \cite{hicks2009ens}
\begin{equation}
M=\frac{\mu_0}{2}\pi a^2 \left ( \frac{1}{R} - \frac{R^2}{(R^2+4 h^2_{eff})^{3/2}} \right ).
\label{eq:susc_mutual}
\end{equation}

The solid line in Fig. \ref{fig:cbcofig2}(b) is obtained by numerically integrating the 2-D Fourier transform of Eq. (\ref{eq:kogsusc}) over the area of the pickup loop, for various values of $z$, and fit to the data by varying $\Lambda$. Figure \ref{fig:cbcofig2}(c) compares the values obtained for the Pearl lengths for a number of the CBCO samples using magnetometry and susceptometry methods. The two methods agree within experimental error over the range of Pearl lengths present. Since the fitting to the Pearl vortex images was done assuming each vortex has $\Phi_0$ of total flux threading through it, rather than a fractional value \cite{mints2000vmc}, this agreement means that the superconducting layers are sufficiently strongly Josephson-coupled that it is energetically favorable for the vortex flux to thread vertically through the superconducting layers, as opposed to escaping between the layers.

\begin{figure}[tb]
        \begin{center}
                \includegraphics[width=10cm]{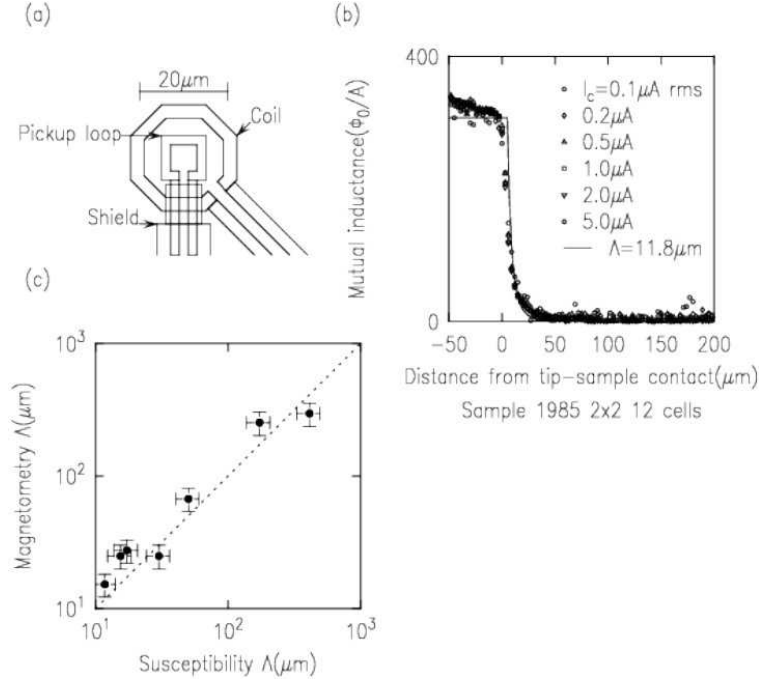}
        \end{center}
        \caption{(a) Geometry of a SQUID susceptometer. (b) Mutual inductance between the field coil and the pickup loop, as a function of the spacing z between the SQUID substrate and a thin CBCO film. The symbols are data, taken with various alternating currents through the field coil. The solid line is modeling using Eq. \ref{eq:thinvort}, with $\Lambda$=11.8$\mu$m. (b) Comparison of the Pearl length $\Lambda$ for a number of CBCO samples using fitting of SQUID magnetometry images of vortices (vertical axis) vs susceptibility measurements (horizontal axis). Reprinted figure with permission from F. Tafuri, J.R. Kirtley, P.G. Medaglia, P. Orgiani, and G. Balestrino,  \href{http://prl.aps.org/abstract/PRL/v92/i15/e157006}{Phys. Rev. Lett. {\bf 92}, 157006 (2004)}. Copyright 2004 by the American Physical Society.}
        \label{fig:cbcofig2}
\end{figure}

Figure \ref{fig:lafepo_susc} shows results from a study using scanning SQUID susceptometry of the penetration depth of the pnictide superconductor LaFePO \cite{hicks2009ens}. This study used the technique of measuring the mutual inductance between the SQUID field coil and pickup loop as a function of spacing $z$, and then repeating the measurement at fixed $z$ while varying the temperature. Since the mutual inductance is a function of $z+\lambda$, the temperature dependence of $\lambda$ can be inferred from such measurements, even without a detailed model for the mutual inductance. Spatially resolved susceptibility measurements (Fig. \ref{fig:lafepo_susc}c-f) showed that the effective height depended on the topography of the sample, with scratches, bumps, and pits effecting the measurements strongly.  However, the measured penetration depth and the temperature dependence of the penetration depth was reproducible for positions more than 10$\mu$m from surface irregularities. In this case the penetration depth had a linear temperature dependence at low temperatures, similar to the cuprate high temperature superconductors, indicative of well formed nodes in the energy gap. It should be noted that in this geometry it is $\lambda_{ab}$, the in-plane penetration depth, that controls the mutual inductance.

\begin{figure}[tb]
        \begin{center}
                \includegraphics[width=14cm]{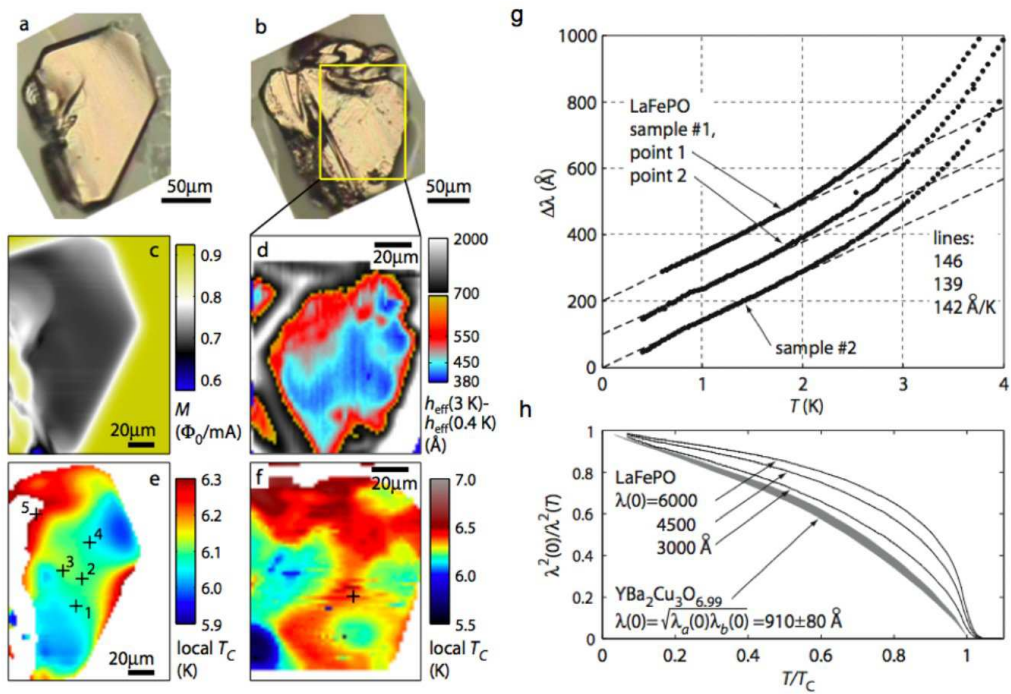}
        \end{center}
        \caption{(a,b) Optical micrographs of two single crystals of the pnictide superconductor LaFePO. (c) Susceptibility scan of \#1 at T=0.4K. (d) Change in $h_{\rm eff}=h+\lambda$ between 0.4 and 3K over sample \#2. (e,f) Maps of local $T_c$ over the same areas as in (c) and (d). The crosses indicate points where $\Delta\lambda(T)$ data were collected. (g) $\Delta\lambda$ of samples \#1, and \#2. The dashed lines are fits between 0.7 K $<$ T $<$ 1.6 K. (h) Black lines are possible superfluid densities for LaFePO sample \#1, point 2, with different $\lambda(0)$. Shaded area: superfluid density of YBa$_2$Cu$_3$O$_{6.99}$ from \cite{bonn2007hhs},\cite{peregbarnea2004avl}. The width of the shaded area reflects uncertainty in $\lambda(0)$. Reprinted figure with permission from C.W. Hicks, T.M. Lippman, M.E. Huber, J.G. Analytis, J.-H. Chu, A.S. Erickson, I.R. Fisher, and K.A. Moler,  \href{http://prl.aps.org/abstract/PRL/v103/i12/e127003}{Phys. Rev. Lett. {\bf 103}, 127003 (2009)}. Copyright 2009 by the American Physical Society.}
        \label{fig:lafepo_susc}
\end{figure}

\begin{figure}[tb]
      \begin{center}
              \includegraphics[width=14cm]{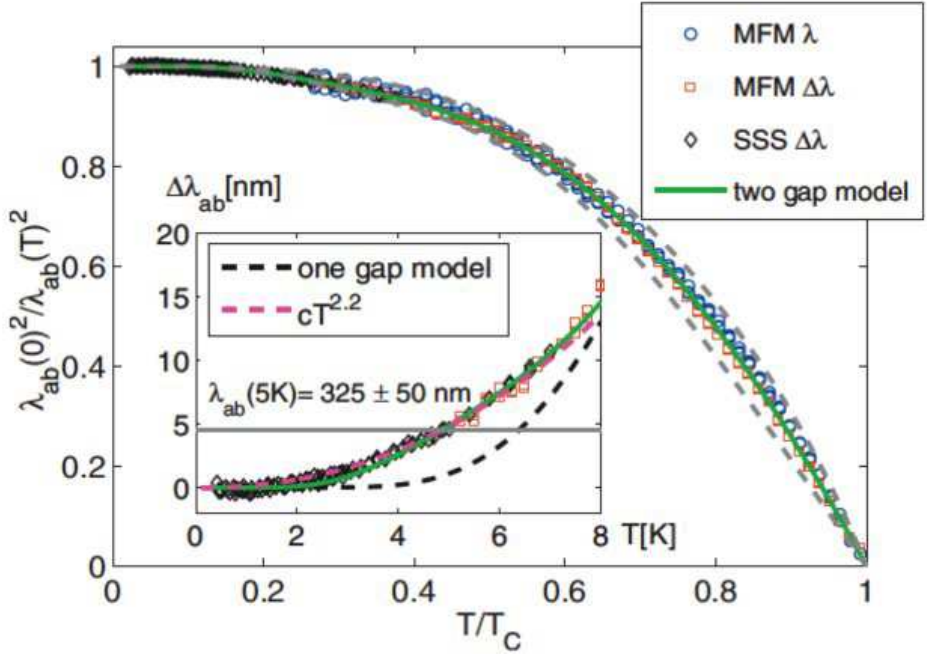}
       \end{center}
       \caption{Normalized superfluid density $\rho_s(T)/\rho_s(0) \equiv \lambda_{ab}(0)^2/\lambda_{ab}(T)^2$ vs $T$ for the pnictide superconductor Ba(Fe$_{0.95}$Co$_{0.05}$)$_2$As$_2$. $\Delta\lambda_{ab}(T)$ is determined from MFM (squares) and SSS (diamonds) experiments by measuring the change in the diamagnetic response at fixed height. These values are offset to match the absolute value of $\lambda_{ab}(T)$ by fitting the MFM data to a truncated cone model (circles). The green solid line shows a fit to a two-band $s$-wave model. The width of the dashed band reflects the uncertainty in $\lambda_{ab}(0)$. Inset: $\Delta\lambda_{ab}(T)$ vs $T$ at low $T$. Black dashed line: one gap $s$-wave model with $a$=1.5 and $\Delta_0=1.95 T_c$. Magenta dashed line: $\Delta\lambda(T)=cT^{2.2}$ (c=0.14 nm/K$^{2.2}$). Reprinted figure with permission from L. Luan, O.M. Auslaender, T.M. Lippman, C.W. Hicks, B. Kalisky, J.-H. Chu, J.G. Analytis, I.R. Fisher, J.R. Kirtley, and K.A. Moler,  \href{http://prb.aps.org/abstract/PRB/v81/i10/e100501}{Phys. Rev. B  {\bf 81}, 100501 (2010)}. Copyright 2010 by the American Physical Society.}
       \label{fig:luan}
\end{figure}  

Perhaps the most promising application of MFM to the study of superconductivity is for high spatial resolution, absolute measurements of penetration depths. There are two ways of inferring superconducting penetration depths from MFM measurements. The first is, as mentioned above, to fit MFM vortex images to a model with the penetration depth as a fitting parameter\cite{nazaretski2009dmp,luan2010lmp}. While Nazaretski {\it et al.} modeled their tip numerically \cite{nazaretski2009dmp}, Luan {\it et al.} used an analytical approach which is easy to fit to experimental data and not strongly affected by differences between tips \cite{luan2010lmp}. A second method is to consider the interaction between the magnetic material in the tip and its image in the superconducting material due to Meissner screening of the tip fields.  Following Xu {\it et al.} \cite{xu1995mlf}, Lu {\it et al.} \cite{lu2002lmp} wrote the force due to the interaction between the magnetic tip, approximated as a magnetic dipole, and its image in the superconductor as
\begin{equation}
F(z)=\frac{3\mu_0m_0^2}{32\pi[z+\lambda(T)]^4},
\label{eq:xmt}
\end{equation}
where $m_0$ is the tip's magnetic moment and $z$ is the physical distance between tip and sample. They fit their measured MFM force derivative curves to the $z>>\lambda$ approximation
\begin{equation}
\frac{dF(z)}{dz}=-\frac{3\mu_0m_0^2}{8\pi[z+\lambda(T)]^5}.
\label{eq:xmt2}
\end{equation}
to obtain the temperature dependence of the penetration depth $\lambda$ for a single crystal of YBCO. Luan {\it et al.} \cite{luan2010lmp} used a truncated cone tip model to write 
\begin{eqnarray}
\left.  \frac{ \partial F_z(z,T)}{\partial z} - \frac{\partial F_z(z,T)}{\partial z}  \right | _{z=\infty}  = \nonumber \\
A \left \{ \frac{1}{z+\lambda_{ab}(T)}+\frac{h_0}{[z+\lambda_{ab}(T)]^2}+\frac{h_0^2}{2[z+\lambda_{ab}(T)]^3} \right \}  
\label{eq:luan}
\end{eqnarray}
for the difference between the force derivative at a given height $z$ and that at large distances, 
where $A$, which depends on the tip shape and coating, is determined from fitting experimental data at low temperatures, $\lambda_{ab}(T)$ is the in-plane component of the penetration depth, and $h_0$ is the truncation height of the tip. Model independent values for $\Delta \lambda(T)$ were obtained by mapping a change in $\lambda$ to a change in $z$ since they are equivalent. Absolute values for $\lambda(T)$ were obtained by fitting to Eq. \ref{eq:luan}. Figure \ref{fig:luan} shows experimental results for the superfluid density $\rho_s$ of the pnictide superconductor Ba(Fe$_{0.95}$Co$_{0.05}$)$_2$As$_2$ obtained by MFM and SSM measurements. The $\Delta \lambda(T)$ results from MFM measurements were later confirmed by a bulk technique on similar samples \cite{kim2010lpd}. The advantage of MFM is that both the absolute value and high sensitivity relative values of $\lambda$ can be obtained in the same measurement, so that superfluid density is obtained over the full temperature range and from it the gap structure of the order parameter. Luan {\it et al.} found a two full-gap model model fit the data well, consistent with the s$^{+-}$ order parameter most frequently discussed in the pnictides. Luan {\it et al.} also found that $\rho_s$ was uniform on the submicron scale despite highly disordered vortex pinning. 

\begin{figure}[tb]
        \begin{center}
                \includegraphics[width=10cm]{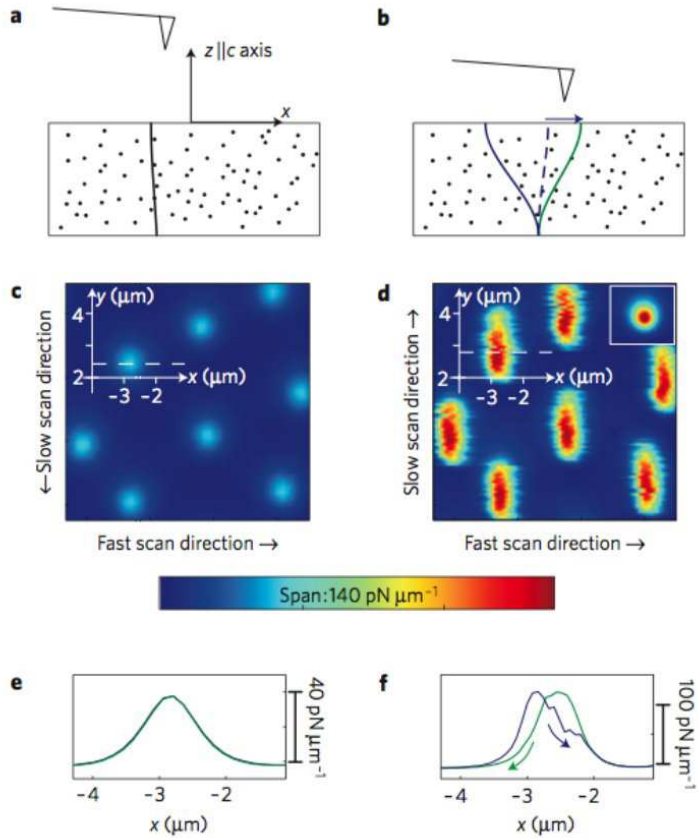}
        \end{center}
        \caption{MFM imaging and manipulation of individual vortices in YBCO at T = 22.3K. {\bf a,b} Schematic drawings of an MFM tip (triangle) that attracts a vortex (thick lines) in a sample with randomly distributed pinning sites (dots): at large heights {\bf a} the force between tip and vortex is too weak to move the vortex; at small heights {\bf b} the vortex moves right and then left as the tip rasters over it. {\bf c} MFM image taken at $z$=420 nm (maximum applied lateral force $F_{max}^{lat} \approx$ 6 nN). {\bf d} $z$=170 nm ($F_{max}^{lat} \approx$ 12 nm). Inset: Scan taken at a comparable height at T=5.2K. {\bf e} Line cut through the data in {\bf c} along the dashed line, showing the signal from a stationary vortex (blue). Overlapping it is a line cut from the reverse scan (green). {\bf f} Line cut through the data in {\bf d} along the dashed line, showing a typical signal from a dragged vortex. Reprinted by permission from Macmillan Publishers Ltd: O.M. Auslaender, L. Luan, E.W.J. Straver, J.E. Hoffman, N.C. Koshnick, E. Zeldov, D.A. Bonn, R. Liang, W.N. Hardy, and K.A. Moler, \href{http://www.nature.com/nphys/journal/v5/n1/abs/nphys1127.html}{Nature Physics {\bf 5}, 35-39 (2008)}. Copyright 2008.}
         \label{fig:auslaender}
\end{figure}

\subsection{Manipulation of individual vortices}
\label{sec:manipulation}
While there is a vast literature on the properties of superconducting vortices \cite{blatter1994vht}, there has been relatively little work on manipulation of individual vortices by scanning magnetic probes. Such manipulation, performed with MFM \cite{hug1994omv,roseman2002mia,moser1998ltm,straver2008cmi,auslaender2009mii,keay2009svh} or SSM \cite{gardner2002msv}, can directly measure the interaction of a moving vortex with the local disorder potential \cite{auslaender2009mii}. Keay {\it et al.}  used MFM to observe sequential vortex hopping between sites in an array of artificial pinning centers \cite{keay2009svh}. An example of another such MFM study is shown in Figure \ref{fig:auslaender} \cite{auslaender2009mii}. Here vortices in a high quality, detwinned sample of YBa$_2$Cu$_3$O$_{6.991}$ are fixed at low temperatures and high scan heights, but at higher temperatures and lower scan heights the tops of the vortices can be dragged by an attractive magnetic interaction with the tip. The vortices are dragged along the fast scan direction until a combination of pinning and elastic forces overcomes the attractive tip-vortex force. The displacement of the vortex in the slow scan direction is much longer than in the fast scan direction, demonstrating a ``vortex wiggling" effect: alternating transverse forces markedly enhance vortex dragging. Vortex motion in both the slow and fast scan directions is stochastic. From such measurements Auslaender {\it et al.} \cite{auslaender2009mii} were able to infer the maximum lateral dragging force before the top of the vortices were depinned. They also found a marked in-plane anisotropy in the distance $w$ the top of the vortex was dragged by the tip in the fast scan direction, larger than could be accounted for by the in-plane anisotropy of the vortex core and point pinning. They speculated that a likely source for extra pinning anisotropy is nanoscale clustering of oxygen vacancies along the Cu-O chains in YBCO.

\section{Conclusions} 

In summary, there are now many types of scanning magnetic microscopies, each with their own strengths and weaknesses. Of the three covered here, SSM is the most sensitive for low resolution applications, but requires a cooled sensor. MFM has the highest spatial resolution, but also applies the largest fields to the sample. SHM can be used over a broader temperature range than SSM. A major thrust in development has been towards higher spatial resolution, which requires smaller sensors scanned closer to the sample. The scaling of resolution and sensitivity with sensor size and spacing depends both on the technique and the source of field. Many of the applications discussed here have involved the imaging of magnetic flux in special geometries, unconventional superconductors, or both. Others have involved time dependence, through {\it e.g.} multiple cooldowns, cooldowns at varying cooling rates, or noise measurements. Some exciting new developments have been scanning susceptibility, which allows the measurement of local penetration depths, spins, and superconducting fluctuations with unprecedented sensitivity and resolution;  MFM imaging of Meissner force derivatives, which allow spatially resolved absolute measurements of penetration depths; and controlled manipulation of superconducting vortices. I expect to see many other exciting developments in the future.

\section{Acknowledgements}

I would like to thank Lan Luan, Beena Kalisky, Cliff Hicks, Danny Hykel and Klaus Hasselbach for carefully reading this review. Any errors are mine, not theirs. I would also like to acknowledge support from the NSF under Grant No. PHY-0425897, by the German Humboldt Foundation, and by the French NanoSciences Fondation, during this work.

\section*{References}
\bibliographystyle{unsrt}
\bibliography{ropsmi_arxiv}

\end{document}